\documentclass[seceq,preprint]{ptptex}
\title{New Parameterization in Muon  Decay\\ 
and the Type of Emitted Neutrino%
\footnote{Submitted to Progress of Theoretical Physics on 
February 15, 2005.}}
\author{Masaru \textsc{Doi},$^{1,}$%
\footnote{E-mail: doi@gly.oups.ac.jp} 
Tsuneyuki \textsc{Kotani}$^{2,}$%
\footnote{E-mail: tsune.kotani@nifty.com} 
and Hiroyuki \textsc{Nishiura}$^{3,}$%
\footnote{E-mail: nishiura@is.oit.ac.jp. 
Present address: Faculty of Information Science and Technology, 
Osaka Institute of Technology, Hirakata 573-0196, Japan.}}

\inst{$^1$Osaka University of Pharmaceutical Sciences, \\
Takatsuki 569-1094, Japan\\
$^2$2-8-26, Higashitoyonaka,
Toyonaka 560-0003, Japan\\
$^3$Department of General Education, \\
Junior College of Osaka Institute of Technology, \\
Osaka 535-8585, Japan}

\recdate{15 February, 2005}

\abst{Normal muon decay, 
$\mu^{+} \to e^{+}\nu_{e}\overline{\nu_{\mu}}$, is studied as 
a tool to discriminate between the Dirac and Majorana types of neutrinos 
and to survey the structure of the weak interaction. 
It is assumed that massive neutrinos mix with one another 
and that the interaction Hamiltonian consists of the $V-A$ and $V+A$ 
charged currents. 
A new set of parameters used in place of the Michel parameters 
is proposed for the positron distribution. 
Explicit forms of these new parameters are obtained 
by assuming that the masses are less than 10 eV for light neutrinos 
and sufficiently large for heavy Majorana neutrinos, 
which are not emitted in the muon decay. 
It is shown that a possible method to discriminate between 
the Dirac and Majorana cases is to use a characterization given 
by the $\chi^2$ fitting of their spectra. 
It is also confirmed that the theoretical predictions 
in the Majorana neutrino case are 
almost the same as those obtained from the standard model. 
Indeed, their differences cannot be distinguished 
within the present experimental precision.
}

\begin{document}
\maketitle
%%%%%%%%%%%%%%%%%%%%%%%%%%%%%%%%%%%%%%%%%%%%%%%%%%%%%%%%%%%%%%%
%%%%%%%%%%%%%%%%%          Section 1         %%%%%%%%%%%%%%%%%%
%%%%%%%%%%%%%%%%%%%%%%%%%%%%%%%%%%%%%%%%%%%%%%%%%%%%%%%%%%%%%%%
\section{Introduction}

The structure of the leptonic charged weak interaction 
provides us with an important source of information 
that may lead to a unified theory beyond the standard model. 
Normal muon decay is a pure leptonic process 
accessible to precise measurements of this structure 
with high statistics, because it is free from the complications 
of the strong interaction and hadronic structure.

Experimental data have been analyzed by employing the 
helicity-preserving four fermion weak interaction with 
$(S \pm P),\,(V \pm A)$ and $T$ 
forms,~\cite{Fetscher}
because this arrangement allows one to make direct contact 
with specific models. The Michel parameters have been used 
to obtain some information concerning the structure of the weak interaction 
under the assumption that the neutrinos are massless
and the lepton number is conserved. Recent experimental data 
have exhibited smaller deviations from the predictions based on 
the standard model~\cite{TwistRho,TwistDelta,TransPol}.

The neutrino emitted in the annihilation of negatively charged leptons 
has been regarded as a particle, while the neutrino created 
together with the negatively charged leptons has been assigned 
to an anti-particle. 
In this assignment, in which there is a distinction between 
the neutrino particle and its anti-particle, 
the neutrino is of the Dirac type, and  
the lepton number is conserved in the weak interaction. 
In the other case, in which there is no such distinction, 
the neutrino is of the Majorana type. 
In the Majorana case, the lepton number is not conserved.

The dominant interaction responsible for muon decay has 
a $V-A$ structure, and the standard model is constructed on this 
footing. In this model, the left-handed neutrino field $\nu_{L}$ 
is assigned to a member of a doublet of the $SU(2)_L\times U(1)$ group, 
and no right-handed field $\nu\,'_{R}$ is present. 
The neutrino is assumed to be massless ($m_{\nu}=0$), and  
the charged current weak interaction takes place 
via the exchange of the left-handed weak gauge boson $W_L$. 
The neutrino and anti-neutrino have the definite helicities 
$h=-1/2$ and $h=+1/2$, respectively, 
and we cannot distinguish between the Dirac and Majorana neutrino 
within the standard model.

Now, however, it has been established through the discovery of neutrino 
oscillation that neutrinos have finite masses and mix with 
one another~\cite{neutrinomass}. 
Thus, it is now known that the 
neutrino cannot be in a definite helicity state. 
For this reason, it is possible to discriminate between neutrinos 
of the Dirac and Majorana types. 
It is an important and fundamental question to determine 
whether the neutrino is of the Dirac or Majorana type. 

There are two other unsolved problems regarding leptons. 
One is to determine why the observed mass differences among the three neutrinos 
are so small in comparison with charged leptons and quarks. 
The other is to understand why the left-handed $V-A$ interaction is favored 
over the right-handed $V+A$ interaction which has not been detected 
definitively.  In the framework of gauge theory, it seems 
natural that the $V-A$ interaction is favored, as a result of the 
spontaneous breakdown of left-right symmetry, which is believed 
to be satisfied at sufficiently high energy.

One appealing way to solve these problems simultaneously 
is to use the idea of the seesaw mechanism, through which the right-handed 
neutrino field $\nu\,'_{R}$ is introduced~\cite{see-saw}. 
Let us explain the scenario of this mechanism 
for the one generation case for simplicity. 
In this scenario, the terms related to the neutrino mass can have 
the Majorana-type mass term $\overline{(\nu\,'_{R})^c}M_{R}\nu\,'_{R}$,
in addition to the Dirac-type mass term, 
$\overline{\nu\,'_{R}}M_D\nu_{L}$.  The Majorana neutrino field 
is defined after diagonalizing the mass matrix,\cite{Doi1} 
(see Eq.~(\ref{eq:Ano03}) in Appendix A). 
The left-handed Majorana neutrino can have a small mass, 
of the order of $M_D^2/M_R$, while the right-handed Majorana 
neutrino can have a large mass, of the order of $M_R$, 
provided that the condition $M_D \ll M_R$ is satisfied. 
This situation leads to various extensions of the standard model, 
such as the $SU(2)_L \times SU(2)_R \times U(1)$ model. 
If this scenario is indeed valid, then we can conclude that 
there exists a right-handed gauge 
boson $W_R$ and that the muon decay receives contributions from interactions 
whose structure is somewhat different from that of the standard model.

It is not yet known whether the neutrino is of the Dirac or Majorana 
type and, further, what structure the weak interaction has beyond the 
standard model. 
For this reason, it is necessary to construct a method 
that provides some information concerning these points. 
Neutrinoless double beta decay 
which violates lepton number conservation is the only 
presently known 
possible way to directly determine the type of the neutrino.  
However, this decay process requires very high resolution 
experimentally, because of the long half-lives, due to 
the tiny neutrino mass and/or the small contribution from the 
$V+A$ current.  Thus, this method is yet to provide 
a decisive conclusion.

Muon decay takes place irrespective of the type of neutrino involved 
and precise data have been accumulated and analyzed 
to investigate the structure of the weak interaction 
by assuming the neutrino to be massless and of the Dirac type.  
It is also meaningful, as a complementary study, to survey 
the possibility of whether muon decay can be used as a 
tool to determine the type of the neutrino.  
With this in mind, we note that there is a difference between
the spectra of emitted positrons in the Dirac and 
Majorana neutrino cases.

The aim of this paper is to propose a new parameterization 
of muon decay that is suitable for analyzing the type of 
neutrinos and the structure of the weak interaction.  
We adopt a Hamiltonian consisting of 
both $V-A$ and $V+A$ currents, which is inspired by the 
$SU(2)_L \times SU(2)_R \times U(1)$ gauge model, 
and present a method to analyze the implications of the 
experimental data.

In \S\ref{sec:two}, we summarize the general framework of 
our study and discuss kinematical effects on the emitted $e^{\pm}$ 
due to finite neutrino mass ($m_{\nu} \neq 0$) . 
There, our assumptions and the approximation adopted in our analysis 
are discussed. 
In \S\ref{sec:three}, the $e^{\pm}$ energy spectrum is surveyed 
in detail. We propose some parameterizations to discriminate between 
the types of neutrinos and discuss their experimental feasibility. 
The polarization of the $e^{\pm}$ is discussed in \S\ref{sec:four}. 
A summary and conclusion are given in \S\ref{sec:five}. 
In Appendix A, to make the paper self-contained, 
features of the lepton mixing matrix for a model 
with left- and right-handed neutrinos are summarized. 
Also there, the details of the coupling 
constants for the weak interaction Hamiltonian based on 
the $SU(2)_L \times SU(2)_R \times U(1)$ gauge model 
are presented for convenience.
In Appendix B, definitions of various coefficients are listed, 
and their explicit forms under certain conditions are given 
for the Dirac and Majorana neutrino cases separately.
%

%%%%%%%%%%%%%%%%%%%%%%%%%%%%%%%%%%%%%%%%%%%%%%%%%%%%%%%%%%%%%%%
%%%%%%%%%%%%%%%%%          Section 2         %%%%%%%%%%%%%%%%%%
%%%%%%%%%%%%%%%%%%%%%%%%%%%%%%%%%%%%%%%%%%%%%%%%%%%%%%%%%%%%%%%
\section{\label{sec:two}General framework}

We assume the following form of the effective weak interaction Hamiltonian 
for the $\mu^{\pm}$ decay:~\cite{Doi1} 
\begin{equation}
{\cal{H}}_W(x)=\frac{G_F}{\sqrt{2}}\{ 
    j_{eL\, \alpha}^\dagger j_{\mu L}^\alpha
    + \lambda j_{eR\,  \alpha}^\dagger j_{\mu R}^\alpha
    + \eta j_{eR\,  \alpha}^\dagger j_{\mu L}^\alpha
    +\kappa j_{eL\,  \alpha}^\dagger j_{\mu R}^\alpha
    \} +\mbox{H.c.}\ ,
    \label{Hamiltonian}
\end{equation}
where $G_{F}$ is the Fermi coupling constant. 
The left-handed and right-handed charged weak leptonic currents, 
$j_{\ell L}$ and $j_{\ell R}$, are defined as
\begin{equation}
j_{\ell L\, \alpha}(x)
  = \overline{{\cal{E}_{\ell}}(x)}\gamma_\alpha(1-\gamma_5)
    \nu_{\ell L}(x)\quad \mbox{and}\quad
j_{\ell R\, \alpha}(x)
  = \overline{{\cal{E}_{\ell}}(x)}\gamma_\alpha(1+\gamma_5)
  \nu_{\ell R}^{\, \prime} (x).
     \label{eq:2no02}
\end{equation}
Here ${\cal{E}}_{\ell}$,  $\nu_{\ell L}$ and $\nu_{\ell R}$ 
are the weak eigenstates of the charged lepton, left-handed neutrinos 
and right-handed neutrinos, with flavors $\ell=e$ and $\mu$.  
The interaction in Eq.~(\ref{Hamiltonian}) is a general form of the 
four fermion, derivative-free Lorentz-invariant interaction, 
which consists of the $V-A$ and $V+A$ currents.

The weak eigenstates of the charged leptons (${\cal{E}}_{\ell}$) and 
neutrinos ($\nu_{\ell L}$ and $\nu_{\ell R}$) are defined, 
respectively, as the superpositions of the mass eigenstate charged 
leptons $E_{\ell}$ and neutrinos $N_j$ with mass $m_j$.  
Then, the charged currents are expressed as
\begin{equation}
\begin{array}{l}
{\displaystyle j_{\ell L\, \alpha}(x)
  = \sum_{j=1}^{2n}\overline{{E_{\ell}}(x)}\gamma_\alpha(1-\gamma_5)
    U_{\ell j}N_j(x),}\\
{\displaystyle j_{\ell R\, \alpha}(x)
  = \sum_{j=1}^{2n}\overline{{E_{\ell}}(x)}\gamma_\alpha(1+\gamma_5)
    V_{\ell j}N_j(x),}
\end{array}
    \label{eq:2no03}
\end{equation} 
for the case of the $n$ generations.~\cite{Doi1}  Here 
$U_{\ell j}$ and $V_{\ell j}$ are, respectively, the left-handed 
and right-handed lepton mixing matrices. 

The weak interaction given in Eq.~(\ref{Hamiltonian}) is naturally 
expected from the gauge models that contain 
the left-handed  and right-handed weak gauge bosons, $W_{L}$ and $W_{R}$. 
In these models, the appearance of the coupling constant 
$\lambda$ is due to $W_{R}$, while terms with $\eta$ and 
$\kappa$ come from 
the possible mixing between $W_{L}$ and $W_{R}$.  As a 
typical example with right-handed interactions, 
we consider the $SU(2)_{L} \times SU(2)_{R} \times U(1)$ model 
in Appendix \ref{sc:AT1}.  If this model is assumed, 
then the coupling constants $\kappa$ and $\eta$ in 
Eq.~(\ref{Hamiltonian}) can be taken as identical, 
as shown in Eq.~(\ref{eq:Ano20}).  However, they are 
treated as independent constants in this paper in order to 
allow comparison with the more general case without a restriction 
from the gauge theory (see, e.g., Ref.~\citen{Fetscher}).  
The structures and magnitudes of the lepton mixing 
matrices $U$ and $V$ are also briefly summarized in Appendix 
\ref{sc:AT1}.  We do not take account of the mirror lepton currents 
and Higgs boson exchange, for simplicity.

Now we study the normal muon decay 
$\mu^{+} \rightarrow e^{+}\nu_e\overline{\nu_\mu}$~
(or $\mu^{-} \rightarrow e^{-}\overline{\nu_e}\nu_\mu$). 
In the framework of the effective weak interaction 
Hamiltonian given in Eq.~(\ref{Hamiltonian}), 
$\mu^{\pm}$ decay takes place as
\begin{equation}
\mu^{\pm} \rightarrow e^{\pm} \ +\ N_j \ +
    \ \overline{N_k},
    \label{eq:2no04} 
\end{equation}
where $\overline{N_k}$ represents an antineutrino for the 
Dirac neutrino case, but it should be understood as 
$N_k$ for the Majorana neutrino case.

If the radiative corrections are not 
included,~\cite{Arbuzov} the differential 
decay rate for polarized $e^{\pm}$ in the 
rest frame of polarized $\mu^{\pm}$ is expressed 
as\cite{Doi2, Fetscher}
\begin{equation}
\frac{d^2 \Gamma(\mu^{\pm} \rightarrow e^{\pm}\nu \overline{\nu})}
     {d x \, d \cos \theta}
     =\left( \frac{m_{\mu} \, G_{F}^2 \, W^4}
     {6 \cdot 4 \, (\pi)^3} \right)
     \, \sqrt{x^2 - x_{0}^2} \, A\,D(x, \, \theta) \, 
      [1 + \vec{P_{e}}(x, \, \theta) 
      \cdot \hat{\zeta} \,\, ],
       \label{eq:2no05}
\end{equation}
where 
\begin{equation}
     x  =  \frac{E}{W} \hspace{5mm} \mbox{and} \hspace{5mm}
     W  = \frac{\, m_{\mu}^{2} + m_{e}^{2} \,}
     {2 \,  m_{\mu}}  = \, 52.8 \,\, \mbox{MeV}.
     \label{eq:2no06}
\end{equation}
Here $m_\mu$ and $m_e$ are the muon and electron masses, respectively, 
and $E$ is the energy of $e^{\pm}$. 
The angle $\theta$ is the direction of the emitted $e^{\pm}$ 
with respect to the muon polarization vector 
$\vec{P_{\mu}}$ at the instant of the $\mu^{\pm}$ decay. 
By taking account of the finite neutrino 
mass ($m_{\nu} \ne 0$),  
the allowed range of $x$ is limited kinematically as 
\begin{eqnarray}
x_{0} \leqq x \leqq x_{\mathrm{max}}=( \, 1 - r_{jk}^{2} \, ),
     \label{eq:2no08}
\end{eqnarray}
where 
\begin{equation}
x_{0} =  \frac{m_{e}}{W} 
     = \, 9.65 \cdot 10^{-3} \quad \mbox{and} \quad 
r_{jk}^{2} =  \frac{(m_{j} + m_{k})^{2}}
     {2 \, m_{\mu} \, W}  = \, 3.63 \cdot 10^{-14}.
     \label{eq:2no09}
\end{equation}
Here $m_j$ and $m_k$ are masses of the neutrinos emitted in the muon decay 
and should satisfy the relation
\begin{equation}
m_j + m_k < (m_\mu - m_e).
   \label{eq:2no10}
\end{equation}
In Eq.~(\ref{eq:2no09}), the neutrino masses have been taken as 
$m_{j}=m_{k}=10\, \mbox{eV}$, 
in order to obtain a rough idea of the magnitude of $r_{jk}^{2}$.

The constant $A$ in Eq.~(\ref{eq:2no05}) is introduced to simplify the 
expression for the differential decay rate 
by using the arbitrariness of its normalization. 
It is referred to as a normalization factor in this paper.  It is shown 
in \S\ref{sec:three} that there are various possibilities for the choice 
of $A$, when experimental data are 
analyzed, although these choices differ only by rearrangements 
of the terms in the theoretical expression.  This constant $A$ plays 
the role of the leading term in the overall normalization 
related to the muon lifetime.

In the differential decay rate given in Eq.~(\ref{eq:2no05}), 
$D(x, \, \theta)$ is the $e^{\pm}$ energy spectrum part, expressed as
\begin{equation}
D(x, \, \theta) = [ \, N(x) 
      \pm P_{\mu} \, \cos \theta \, P(x) \, ],
      \label{eq:2no07}
\end{equation}
where $P_{\mu}=|\vec{P_{\mu}}|$, and the functions $N(x)$ and 
$P(x)$ are, respectively, the isotropic and anisotropic parts of the 
$e^{\pm}$ energy spectrum.  Their details are discussed in 
\S\ref{sec:three}. The plus 
(minus) sign in Eq.~(\ref{eq:2no07}) corresponds to 
$\mu^{+}$ ($\mu^{-}$) decay.  The vector 
$\vec{P_{e}}(x, \, \theta)$ in Eq.~(\ref{eq:2no05}) is a 
polarization vector of $e^{\pm}$, and $\hat{\zeta}$ is the 
directional vector of the measurement of the $e^{\pm}$ spin 
polarization.  We discuss $\vec{P_{e}}(x, \, \theta)$ 
in \S\ref{sec:four}.

The isotropic and anisotropic parts of the energy spectrum 
consist of various terms, 
each of which has a different $x$ dependence taking a complicated 
form and includes some combination of the lepton mixing matrices. 
Therefore, in order to extract useful information, it is 
desirable, as the first step, to investigate the characteristic 
features of their $x$ dependences and to simplify them. 

As an example, let us consider only two terms of $N(x)$,
\begin{eqnarray}
N(x) = \left( \frac{1}{A} \right)
\left[ a_{+} (3 x - 2 x^{2} - x_{0}^{2})
     + (b_{+} - a_{+}) \, (2 x - x^2 - x_{0}^{2}) \right] 
     ,
     \label{eq:2no11}
\end{eqnarray}
where
\begin{align}
&a_{+} = ( a + \lambda^2 \, \hat{a} ),&
&b_{+} = ( b + \lambda^2 \, \hat{b} ),
     \label{eq:2no12} \\
&a = {\mathrm{\Sigma}}_{jk}^{'} \, F_{j \, k}^{3/2}
     |U_{ej}|^{2} |U_{\mu k}|^{2},&
&b = {\mathrm{\Sigma}}_{jk}^{'} \, F_{j \, k}^{1/2} G_{j \, k} \,
     |U_{ej}|^{2} |U_{\mu k}|^{2},
     \label{eq:2no13a} \\
&\hat{a} = {\mathrm{\Sigma}}_{jk}^{'} \, F_{j \, k}^{3/2}
     |V_{ej}|^{2} |V_{\mu k}|^{2},&
&\hat{b} = {\mathrm{\Sigma}}_{jk}^{'} \, F_{j \, k}^{1/2} G_{j \, k} \,
     |V_{ej}|^{2} |V_{\mu k}|^{2}.
     \label{eq:2no13b}
\end{align}
Here, the primed sum represents the sum taken over neutrinos 
whose emission in the $\mu^{\pm}$ decay is allowed 
by the restriction given in Eq.~(\ref{eq:2no10}).

Two kinematical factors, $F_{j\,k}$ and $G_{j\,k}$, emerge from 
the phase space integral of the emitted neutrinos in both the 
Dirac and Majorana neutrino cases.  
They represent the additional kinematical effect 
due to the finite mass of the neutrino ($m_{\nu} \neq 0$), 
and their explicit forms are given in 
\S\ref{sec:2no1} [see Eqs.~(\ref{eq:2no14}) and (\ref{eq:2no15})].  
The treatment of the lepton mixing 
matrices, $U_{\ell \, j}$ and $V_{\ell \, j}$, 
is considered in \S\ref{sec:2no2}.

\subsection{\label{sec:2no1}Kinematical factors due to 
the non-vanishing neutrino mass $\,\cdots \,$ {\rm Condition (A)}}

First, it is worthwhile to note that all terms in the spectrum 
$D(x, \, \theta)$ and the polarization 
$\vec{P_{e}}(x, \, \theta)$ in Eq.~(\ref{eq:2no05}) 
are proportional to $F_{j\,k}^{1/2}$, for example, as shown in 
Eq.~(\ref{eq:2no13a}).  These terms include some 
combinations of the following kinematical factors:%
%%%%%%%%%%%%%%%%%%% footnote begin  %%%%%%%%%%%%%%%%%%
\footnote{Strictly speaking, some terms proportional to 
$m_{\nu}$ include another kinematical factor accompanied 
by $F_{j\,k}^{1/2}$ (see Ref.~\citen{Doi3}).  However, 
we do not consider these terms in this paper, 
because their contributions are negligible.}
%%%%%%%%%%%%%%%%%%  footnote end  %%%%%%%%%%%%%%%%%%%%%%%%
\begin{eqnarray}
F_{j\,k}
  &=& \left[ 1 - \left( \frac{r_{j\,k}^2}{1-x} \right) \right]
     \left[ 1 - \left( \frac{r_{j\,k}^2}{1-x} \right)
     \left( 1 - 4 \, \mu_{j\,k} \right) \right],
     \label{eq:2no14} \\
G_{j\,k}
  &=& \left[ 1 - \left( \frac{r_{j\,k}^2}{1-x} \right) \right]
     \left[ 1 + 2 \left( \frac{r_{j\,k}^2}{1-x} \right)
     \left( 1 - \mu_{j\,k} \right) \right]
     + 6 \, \left( \frac{r_{j\,k}^2}{1-x} \right)^{2}\mu_{j\,k},
     \label{eq:2no15}
\end{eqnarray}
where 
$\mu_{j\,k}$ is defined by
\begin{eqnarray}
\mu_{j\,k} =  \frac{m_{j} \, m_{k}}
     {(m_{j} + m_{k})^{2}}. 
    \label{eq:2no16}
\end{eqnarray}

If $m_{\nu} = 0$, both $F_{jk}$ and 
$G_{jk}$ are unity for the entire range of $x$, as seen 
from Eqs.~(\ref{eq:2no14}) and (\ref{eq:2no15}).
However, if $m_{\nu} \ne 0$, 
they display the following different types of behavior near $x_{\mathrm{max}}$: 
\begin{eqnarray}
F_{j\,k} \to 0 \hspace{5mm} \mbox{and}
     \hspace{5mm} G_{j\,k} \to 6\mu_{j\,k}
     \hspace{5mm} \mbox{in the limit} \,\,
     x \to x_{\mathrm{max}}=( \, 1 - r_{jk}^{2} \, ) .
     \label{eq:2no17}
\end{eqnarray}
This implies that the spectrum and polarization of $e^{\pm}$ 
tend to zero suddenly near $x_{\mathrm{max}}$.%
%%%%%%%%%%%%%  footnote begin  %%%%%%%%%%%%%%%%%%%%%
\footnote{Here we state the reason why $F_{jk}$ suddenly becomes 
zero near $x_{\mathrm{max}}$ for $m_{\nu} \ne 0$, in 
spite of the fact that $F_{jk} = 1$ for $m_{\nu} = 0$.  It is convenient to 
introduce the momentum transfer squared, $\Delta^{2} = 
(q_{j} + q_{k})^{2}$, where $q_{j}$ is the 4-dimensional 
momentum of $N_{j}$.  
Because the $e^{\pm}$ energy $E$ is given by 
$E=(m_\mu^2+m_e^2-\Delta^2)/2m_\mu$, 
the maximum energy is realized when $\Delta^{2}$ 
takes the minimum value $\Delta^{2} = (m_{j} + m_{k})^{2}$, 
as shown in Eq.~(\ref{eq:2no08}).  
This minimum is realized under the following 
two conditions: ($\mathrm{i}$) Both of the neutrinos are emitted in 
the direction opposite to $e^{\pm}$ in order to satisfy the 
momentum conservation, namely, 
$\vec{q_{j}}+\vec{q_{k}}+\vec{p_{e}} = 0$.  ($\mathrm{ii}$) 
Each neutrino has a definite momentum, say 
$q_{j} = p_{e} \, m_{j} / (m_{j} + m_{k})$.  Therefore, there is 
no freedom to assign arbitrary neutrino momentum.  In other words, 
the density in the phase space is zero; that is, we have 
$F_{jk} = 0$.  A similar situation exists in the case that 
the mass of one neutrino is zero.  Contrastingly, if 
$m_{j} = m_{k} = 0$, we have to choose $\Delta^{2} = 0$. 
This situation is allowed for various 
combinations of two neutrino momenta, 
since only the total momentum of the neutrinos is fixed 
under the condition ($\mathrm{i}$).  
Thus, there is no special restriction in 
the phase space compared to the general case for 
an arbitrary $e^{\pm}$ energy; that is, we have $F_{jk} = 1$.}
%%%%%%%%%%%%%  footnote end  %%%%%%%%%%%%%%%%%%%%%%%%%%%%%
Therefore, in principle, the shape of the spectrum near 
$x_{\mathrm{max}}$ exhibits different behavior 
for the massive and massless neutrino cases. 

Fortunately, the $x$ dependences of 
$F_{jk}$ and $G_{jk}$ appear significantly only in 
a very tiny range near $x_{\mathrm{max}}$, 
say, for $x >(1-10^8r_{jk}^{2})\sim(1-10^{-6})$ 
if the required numerical accuracy of the experiment is of 
order $10^{-6}$.  Thus, practically, there seems to be no problem 
in treating $F_{jk}$ and $G_{jk}$ in $D(x, \, \theta)$ and 
$\vec{P_{e}}(x, \, \theta)$ as independent of $x$.  
More specifically, if we assume that 
the emitted neutrinos have small masses, 
i.e., at most of the order of $10 \, \mbox{eV}$,  
the following approximations yield very good accuracy:
\begin{eqnarray} 
F_{j\,k} = 1 \hspace{5mm} \mbox{and} \hspace{5mm}
     G_{j\,k} = 1.
     \label{eq:2no18}
\end{eqnarray}
Hereafter, this approximation is referred to as Condition (A).

Under Condition (A), the second term of $N(x)$ in 
Eq.~(\ref{eq:2no11}) yields no contribution.  Hence, we need not consider 
this term, except for the negligible tiny range near $x_{\mathrm{max}}$.  
We omit such terms in this paper without any loss of practical 
accuracy.

\subsection{\label{sec:2no2}Contributions from the lepton mixing 
matrices U and V $\cdots$ {\rm Condition (B)}}

The sum over the square (or product) of the lepton mixing matrix 
elements appears in the spectrum $D(x, \, \theta)$ and 
polarization $\vec{P_{e}}(x, \, \theta)$ appearing in Eq.~(\ref{eq:2no05}), 
for example as seen from  Eqs.~(\ref{eq:2no13a}) and (\ref{eq:2no13b}).  
If Condition (A) is accepted, we are able to take the 
sum over neutrino indices under some assumptions.

In the Dirac neutrino case, it is assumed that all neutrinos can 
be emitted in the $\mu^{\pm}$ decay. 
Then we have the following properties from the unitarity conditions 
of $U$ and $V$:
\begin{equation}
{\mathrm{\Sigma}}_{j}|U_{\ell j}|^2 = {\mathrm{\Sigma}}_{j}|V_{\ell j}|^2=1.
\label{eq:2no19}
\end{equation}

By contrast, in the Majorana neutrino case, 
we assume the existence of heavy Majorana neutrinos, 
which are not emitted in the $\mu^{\pm}$ decay.   
Then, there is a different situation, in which we have
\begin{equation}
{\mathrm{\Sigma}}_{j}^{ \, \prime}|U_{\ell j}|^2  = 1-\overline{u_{\ell}}^{\, 2}
\quad \mbox{and} \quad
{\mathrm{\Sigma}}_{j}^{ \, \prime}|V_{\ell j}|^2  = \overline{v_{\ell}}^{\, 2},
\label{eq:2no21}
\end{equation}
where the primed sums are taken over only the light 
neutrinos.  Here $\overline{u_{\ell}}^{\, 2}$ and 
$\overline{v_{\ell}}^{\, 2}$ 
are the representatives of small deviations from unitarity due 
to these heavy Majorana neutrinos.  [ The details are given in 
Eqs.~(\ref{eq:Ano07})~--~(\ref{eq:Ano09}),  
(\ref{eq:Ano14}) and (\ref{eq:Ano15}).]
In what follows, the explicit forms of these matrices $U$ and $V$ 
are not needed.

In addition, in the Majorana neutrino case, the following products 
of $U$ and $V$ appear:
\begin{equation}
\overline{w_{e \mu}}
  \equiv {\mathrm{\Sigma}}_{j}^{\, \prime} \, U_{ej} \, V_{\mu j}
     \quad \mbox{and} \quad
\overline{w_{e \mu \, h}}
  \equiv {\mathrm{\Sigma}}_{k}^{\, \prime} \, V_{ek} \, U_{\mu k}.
    \label{eq:2no23}
\end{equation}
An example in which $\overline{w_{e \mu}}$ appears is mentioned 
in \S\ref{sec:2no3}.  The quantities $\overline{w_{e \mu}}$ and 
$\overline{w_{e \mu \, h}}$ are also small, 
as shown in Appendix~\ref{sc:AT1} and Eq.~(\ref{eq:4no13}).  
The assumptions Eqs.~(\ref{eq:2no19})~--~(\ref{eq:2no23}) are 
referred to as Condition (B).

\subsection{\label{sec:2no3}Terms characteristic of the 
Majorana-type neutrino}

Finally, let us explain why the Majorana-type neutrino 
offers different information from the Dirac-type 
neutrino.  We emphasize that some contributions that are specific 
to the Majorana neutrino case are added to the decay rate.  Among 
these, one appears even in the $(V-A)$ interaction 
alone if $m_{\nu}\ne 0$.  The others appear if the $(V+A)$ 
interaction exists in addition 
to $(V-A)$, even in the case $m_{\nu} = 0$.  
We study two examples.

As the first example, let us consider 
the mode in which both the ($e^{+}$, $N_{j}$) and 
($\mu^{+}$, $\overline{N_{k}}$) vertices are of the 
$(V-A)$ type.  This assignment is referred to as the ordinary 
mode (A).  If $m_{\nu}\ne 0$, $N_{j}$ has a small component of 
helicity $h = + 1/2$, whose magnitude is proportional to 
$m_{j} / \omega_{j}$, where $\omega_{j}$ is the energy of $N_{j}$.  
Therefore, in the Majorana neutrino case, in which 
this $N_{j}$ cannot be distinguished from the
$\overline{N_{j}}$ associated with $\mu^{+}$, 
there is a new cross mode (B) for which $N_j~(=\overline{N_j})$ comes 
from the annihilation of $\mu^+$ and $\overline{N_k}~(=N_k)$ is 
emitted with $e^{+}$.  The interference between the ordinary 
mode (A) and the cross mode (B) appears in the 
decay probability.  It becomes proportional to the 
product of the neutrino masses, $m_j \, m_k$.  Of course, 
there is no such possibility in the Dirac neutrino 
case, where $N_{j} \ne \overline{N_{j}}$.  

Now, if the $(V+A)$ interaction is added to $(V-A)$, 
the situation changes greatly.  As a second 
example, let us consider 
the mode for which the ($e^{+}$, $N_{j}$) vertex is of the 
$(V-A)$ type, but the ($\mu^{+}$, $\overline{N_{k}}$) 
vertex is of the $(V+A)$ type.  In this case, the helicity 
of $\overline{N_{k}}$ is $h = - 1/2$, the same as that 
of $N_{j}$, even in the limit $m_{\nu} \to 0$.  We 
call this case the mode (C).  In the Majorana neutrino case, 
it is possible to have another 
cross emitting mode (D) with the ($e^{+}$, $N_{k}$) and 
($\mu^{+}$, $N_{j}$) vertices, in addition to the mode (C). There arises 
interference between the (C) and (D) modes in the decay 
probability.  The leading term of this interference 
is not proportional to $m_{\nu}$, but it includes the coupling 
constant $\kappa^2$ with lepton mixing matrices 
$|{\mathrm{\Sigma}}_{j}^{\, \prime} \, U_{ej} \, V_{\mu j}|^2$. 
This results from the equality of the helicities, 
mentioned above.  The mode (D) does not exist in the Dirac 
neutrino case. [For details, see Fig.~(2) and 
Table~I of Ref.~\citen{Doi2}]. 

Many terms proportional to $m_{\nu}$ exist in the decay 
rate in both the Dirac and Majorana neutrino cases,  
because of the small component of the helicity that is proportional 
to $m_{\nu}$.  However, these terms are negligibly small and are not 
taken into account in this paper.  The complete decay formulae 
including these terms are given in Refs.~\citen{Doi3} and \citen{Shrock}. 

%%%%%%%%%%%%%%%%%%%%%%%%%%%%%%%%%%%%%%%%%%%%%%%%%%%%%%%%%%%%%%%
%%%%%%%%%%%%%%%%%          Section 3         %%%%%%%%%%%%%%%%%%
%%%%%%%%%%%%%%%%%%%%%%%%%%%%%%%%%%%%%%%%%%%%%%%%%%%%%%%%%%%%%%%
\section{\label{sec:three}Energy spectrum of $e^{\pm}$}

The isotropic part $N(x)$ and anisotropic part $P(x)$
of the $e^{\pm}$ energy spectrum appearing in Eq.~(\ref{eq:2no07}) are, 
respectively, expressed as follows under Conditions (A) and 
(B) given in \S\ref{sec:two}:
\begin{eqnarray}
N(x)
  &=& \left( \frac{1}{A} \right)
  \left[ a_{+} (3 x - 2 x^{2} - x_{0}^{2})
     + 12 \, ( \, k_{+ \, c} + \varepsilon_{m} \, k_{+ \, m} \, )
     \, x \, (1 - x) \right. \nonumber \\
   & & {} \hspace{15mm} \left.
     + 6 \, \varepsilon_{m} \, \lambda \, d_{r} \, 
     x_{0} \, (1-x) \, \right] ,
     \label{eq:3no01} \\
P(x)
 &=& \left( \frac{1}{A} \right) \sqrt{x^{2} - x_{0}^{2}} \,
     \left[ a_{-} (-1 + 2 \, x - r_{0}^{2})
     \right. \nonumber \\
  & & {} \hspace{15mm} \left.
     + 12 \, ( \, k_{- \, c} + \varepsilon_{m} k_{- \, m} \, )
     \, (1 - x) \right] .
     \label{eq:3no02}
\end{eqnarray}
Here, the decay formulae for the Dirac and Majorana
neutrinos are obtained by setting $\varepsilon_{m} = 0$
and $\varepsilon_{m} = 1$, respectively.  The original forms of 
$N(x)$ and $P(x)$ before adopting Conditions (A) 
and (B) are presented in Eqs.~(\ref{eq:Bno06}) and 
(\ref{eq:Bno09}) of Appendix B.

The first terms of $N(x)$ and $P(x)$ represent 
their $x$ dependences obtained from the standard model, 
where $A = a_{\pm} = 1$, and all other coefficients 
($k_{\pm \, c}$, $k_{\pm \, m}$ and $d_{r}$) are 
zero.  The quantity $r_{0}^{2}$ in $P(x)$ is defined as follows:
\begin{eqnarray}
r_{0}^{2} \equiv  \frac{m_{e}^{2}}{m_{\mu} \, W} =
     \left( 1 - \sqrt{1-x_{0}^{2}} \right)
     = 4.66 \cdot 10^{-5} \, .
    \label{eq:3no03}
\end{eqnarray}

Before listing the details of these coefficients, we 
note here that the suffix $c$ of $k_{\pm}$ indicates  
contributions that are the same for the Dirac and Majorana neutrino cases, 
whereas the suffix $m$ indicates coefficients associated with 
the Majorana neutrino only. 

In the Dirac neutrino case, these coefficients are expressed as 
follows:
\begin{eqnarray}
a_{\pm} = \left( 1 \pm \lambda^2 \right)
     \hspace{3mm} \mbox{and} \hspace{3mm}
k_{\pm \, c} = \left( \frac{1}{2} \right)
     \left( \kappa^2 \pm \eta^{2} \right).
     \label{eq:3no14}
\end{eqnarray}
Of course, in this case there are no contributions from $k_{\pm \, m}$ 
and $d_{r}$.  Note that the final expressions for $N(x)$ and 
$P(x)$ in the Dirac neutrino case are the same as those 
obtained by assuming $m_{\nu} = 0$ for one generation if 
Conditions (A) and (B) in \S\ref{sec:two} are both satisfied. 

In the Majorana neutrino case, these coefficients have the following 
complicated forms:
\begin{eqnarray}
a_{\pm} &=&
     \left[ \left( 1 - \overline{u_{e}}^{\, 2} \right)
     \left( 1 - \overline{u_{\mu}}^{\, 2} \right)
     \pm \lambda^2 \, \overline{v_{e}}^{\, 2} \,
     \overline{v_{\mu}}^{\, 2} \right] ,
     \label{eq:3no16} \\
k_{\pm \, c} &=& \left( \frac{1}{2} \right)
     \left[ \kappa^2 \, (1 - \overline{u_{e}}^{\, 2} )
      \, \overline{v_{\mu}}^{\, 2}
      \pm \eta^{2} \, \overline{v_{e}}^{\, 2} \,
      (1 - \overline{u_{\mu}}^{\, 2} ) \right] ,
     \label{eq:3no17} \\
k_{\pm \, m} &=& \left( \frac{1}{2} \right)
     \left[ \,\kappa^2 \,
     | \, \overline{w_{e \mu}} \, |^{2}
     \pm \eta^{2} \,
     | \, \overline{w_{e \mu \, h}} \, |^{2} \, \right] \, ,
     \label{eq:3no18} \\
d_{r} &=& \left( \frac{1}{2} \right)
     \mbox{Re} ( \overline{w_{e \mu}}^{\,\, *} \,\,
     \overline{w_{e \mu \, h}} ).
     \label{eq:3no19}
\end{eqnarray}
Here, $\overline{u_{\ell}}^{\, 2}$, $\overline{v_{\ell}}^{\, 2}$, 
$\overline{w_{e  \, \mu}}$ and $\overline{w_{e \, \mu \, h}}$ 
are small quantities in general, as mentioned with regard to 
Eqs.~(\ref{eq:2no21}) and (\ref{eq:2no23}).  It should be 
noted that $k_{\pm \, c}$ has the same order of 
magnitude as $k_{\pm \, m}$, in contrast to the Dirac 
neutrino case.  It is worthwhile noting that the final 
$N(x)$ and $P(x)$ in the Majorana neutrino case exhibit 
small deviations from those of standard model independently 
of the magnitudes of the coupling constants $\lambda$, 
$\kappa$ and $\eta$.

Since 1985, experimental data have been analyzed using 
expressions based on the well-known Michel parameters.  
Expressing $N(x)$ and $P(x)$ in terms of 
the Michel parameters using our notation, we have 
\begin{eqnarray}
N(x)
  &=& 6 
     \Bigl[  x (1 - x) + \frac{2}{9}\,\rho_{M} \,
     \left( 4 x^{2} - 3 x - x_{0}^{2} \right)
     + \eta_{M} \, x_{0} \, (1-x) \Bigr],
     \label{eq:3no04} \\
P(x) 
  &=& 2 \, \xi_{M} \sqrt{x^{2} - x_{0}^{2}}
      \Bigl[ ( 1 - x )+ \frac{2}{3} \, \delta_{M}
      \, \left( 4 x - 3 - r_{0}^{2} \right) \Bigr].
\label{eq:3no05}
\end{eqnarray}
These expressions are obtained, for example, from Ref.~\citen{Fetscher} 
or Eqs.~(31) and (32) of Ref.~\citen{Kuno}.%
%%%%%%%%%%%%%%%%%%%%%%%%% footnote begin %%%%%%%%%%%%%%%%%%%%%
\footnote{Our definitions and 
those in Ref.~\citen{Fetscher} are related as $N(x) = 6 F_{IS}(x)$ 
and $P(x) = 6 F_{AS}(x)$.  The amplitudes $g_{\varepsilon \mu}^{\gamma}$ 
defined in Ref.~\citen{Fetscher} correspond to our coupling constants 
as follows:
$g_{LL}^{V}=1$, \, $g_{RR}^{V}=\lambda$, \, $g_{LR}^{V}=\kappa$, 
\, $g_{RL}^{V}=\eta$, with all other amplitudes set to zero
in the present paper.}
%%%%%%%%%%%%%%%  footnote end  %%%%%%%%%%%%%%%%%%%%%%%%%%%%%
Hereafter, these forms are referred to as 
the Michel parameterization.  They are presented for the 
Dirac neutrino case with $m_{\nu} = 0$.  In the standard model, 
these parameters take definite values: $\rho_{M} = \delta_{M} = 0.75, 
\, \xi_{M} = 1$, and $\eta_{M} = 0$. In 
our model, there is no $\eta_{M}$ term, 
even in the massive Dirac neutrino case, if we ignore 
terms proportional to $m_{\nu}$.  The reason why 
the $\eta_{M} \ne 0$ term appears in the Michel 
parameterization of Ref.~\citen{Fetscher} is that it comes 
from the interference between the $(V \pm A)$ and 
$(S \pm P)$ (or $T$) interactions.

At first glance, the relation between $N(x)$ in Eq.~(\ref{eq:3no01}) 
and that in Eq.~(\ref{eq:3no04}) may not be clear.  It is 
shown in \S\ref{sec:3no1} that the Michel parameterization given 
in Eq.~(\ref{eq:3no04}) is a special case of our expression appearing in 
Eq.~(\ref{eq:3no01}) in which some appropriate constant value is chosen 
for the normalization factor $A$.  A similar situation is 
demonstrated for $P(x)$ in \S\ref{sec:3no2}.

\subsection{\label{sec:3no1}Isotropic part of the spectrum: $N(x)$}

First, let us introduce the following quantity 
for the normalization factor $A$:
\begin{eqnarray}
A_{n \, \ell} = ( a_{+} + 2 n \, k_{+ \, c}
      + 2 \ell \, \varepsilon_{m} k_{+ \, m} \, ) \, > 0 \, ,
\label{eq:3no06}
\end{eqnarray}
\noindent
where $n$ and $\ell$ are some integers.
Next, we verify that Eq.~(\ref{eq:3no04}) is a special case 
of Eq.~(\ref{eq:3no01}) with the choice of either 
$A = A_{1 \, 0}$ or $A_{1 \, 1}$.  

We first consider the case 
with $A = A_{1 \, 0}$, in which terms characteristic of the 
Majorana neutrino case appear explicitly.  Then, the following 
expression is derived:
\begin{eqnarray}
N(x) 
  &=& \left( \frac{1}{A_{1 \, 0}} \right)
     \Bigl\{ ( a_{+} + 2 \, k_{+ \, c})
     (3 x - 2 x^{2} - x_{0}^{2}) 
     \Bigr. \nonumber \\
   & & {} \hspace{23mm} \Bigl. + k_{+ \, c}
     [ 12 \, x \, (1 - x) - 2 (3 x - 2 x^{2} - x_{0}^{2})]
     \Bigr. \nonumber \\
   & & {} \hspace{23mm} \Bigl. + \varepsilon_{m} \,
     \left[ 12 \, k_{+ \, m} \, x \, ( 1 - x )
     + 6 \lambda \, d_{r} \, x_{0} \, ( 1 - x ) \right]
     \Bigr\} 
     \label{eq:3no07} \\
  &=& \Bigl[ (3 x - 2 x^{2} - x_{0}^{2}) 
     + 2 \, \rho_{c}
      \left( 3 x - 4 x^{2} + x_{0}^{2} \right)
      \Bigr. \nonumber \\
   & & {} \hspace{23mm} \Bigl. + 12 \, \varepsilon_{m} \,
      \rho_{m} \, x (1 - x) + 6 \, \varepsilon_{m} \,
      \eta_{m} \, x_{0} \, ( 1 - x ) \Bigr].
\label{eq:3no08}
\end{eqnarray}
\noindent
Here, the parameters are defined as follows:
\begin{eqnarray}
\rho_{c} = \left( \frac{k_{+ \, c}}{A_{1 \, 0}} \right)
     > 0, \hspace{3mm} \rho_{m} =
     \left( \frac{k_{+ \, m}}{A_{1 \, 0}} \right) > 0
     \hspace{3mm} \mbox{and} \hspace{3mm}
     \eta_{m} &=&
      \left( \frac{\lambda \, d_{r}}{A_{1 \, 0}} \right).
\label{eq:3no09}
\end{eqnarray}
\noindent
The fact that $\rho_{c}$ and $\rho_{m}$ are positive 
is clear from Eqs.~(\ref{eq:3no14})~--~(\ref{eq:3no18}) and 
(\ref{eq:3no06}).
Now, it is easy to confirm that the Michel parameterization, 
Eq.~(\ref{eq:3no04}), can be obtained from 
Eq.~(\ref{eq:3no08}) by introducing the relation
\begin{eqnarray}
     2 \, \rho_{c} = \left( 1 - \frac{4}{3} \, \rho_{M} \right),
     \label{eq:3no12}
\end{eqnarray}
\noindent
because, for this confirmation, it is not necessary to take account 
of the terms $\rho_{m}$ and $\eta_{m}$ appearing only in the 
Majorana neutrino case.  Note that the relation $A = 1$ is 
required in the Michel parameterization 
[see Eq.~(44) of Ref.~\citen{Kuno}].

Next, if the case with $A = A_{1 \, 1}$ is chosen, we again 
obtain the form given in Eq.~(\ref{eq:3no04}) if we replace 
$\rho_{c}$ in Eqs.~(\ref{eq:3no08}) and (\ref{eq:3no12}) by 
$(\rho_{c} + \varepsilon_{m} \rho_{m})$, because the third term 
in Eq.~(\ref{eq:3no08}) is absorbed into the second term 
through this replacement, and the fourth term 
plays the role of the $\eta_{M}$ 
parameter.  Of course, the denominator $A_{1 \, 0}$ 
of Eq.~(\ref{eq:3no09}) should be replaced 
by $A_{1 \, 1}$.

Thus, it is clear that there is no advantage 
of introducing the Michel parameter at present, although 
it was very useful in determining the type of the weak 
interaction.  In fact, in the Michel parameterization 
defined in Ref.~\citen{Fetscher}, 
the form $(\rho_{M} - 3/4)$, like Eq.~(\ref{eq:3no12}) [and 
$(\delta_{M} - 3/4)$, like Eq.~(\ref{eq:3no26})] is used.  
Therefore, instead of measuring the deviation from 
$\rho_{M} = 0.75$, it seems desirable to 
directly determine $\rho_{c}$ and $\rho_{m}$ 
themselves, which indicate the deviations from the
standard model.
Furthermore, it is worthwhile to note that we are interested in 
estimating the $x$ dependence of the difference between the 
experimental data and the prediction of the standard model, 
namely, $(3x-2x^2-x_0^2)$ appearing in Eq.~(\ref{eq:3no01}), 
through use of the $\chi^2$-fitting in the experimental analysis.

Recently, the TWIST group reported 
a precise experimental result for $\rho_{M}$,\cite{TwistRho}
\begin{eqnarray}
  \rho_{M} = 0.75080 \pm 0.00032\mathrm{(stat)} \pm
   0.00097\mathrm{(syst)} \pm 0.00023,
     \label{eq:3no20}
\end{eqnarray}
\noindent
where the third error comes from the ambiguity in $\eta_{M}$ 
appearing in Eq.~(\ref{eq:3no04}).  
This ambiguity is due to the fact that various values for $\eta_{M}$ 
have been used within the uncertainty of the accepted average 
value $\eta_{M} = (-7 \pm 13) \cdot 10^{-3}$.~\cite{Fetscher}
Assuming the Dirac-type neutrino, this group obtained the result 
$|\tan \zeta | < 0.030$ by combining Eqs.~(\ref{eq:3no12}), 
(\ref{eq:3no09}) and (\ref{eq:3no14}) with Eq.~(\ref{eq:Ano20}), where 
$\zeta$ is the $W_{L}$\,-\,$W_{R}$ mixing angle defined in 
Eqs.~(\ref{eq:Ano16}) and (\ref{eq:Ano17}).

Although the TWIST group analyzed their data using 
Eq.~(\ref{eq:3no04}), their results can be interpreted in 
our analysis as follows.  If neutrinos are of the Dirac type, there 
is no difference between the choices $A = A_{1 \, 0}$ 
and $A = A_{1 \, 1}$, because in this case $\rho_{m} = \eta_{m} = 0$. 
For the Majorana neutrino case, we choose $A = A_{1 \, 1}$.  
Then, the restriction on $(\rho_{c} + \varepsilon_{m} \rho_{m})$ 
can be obtained from their result for $\rho_{M}$ by using 
the relation in Eq.~(\ref{eq:3no12}), 
and some information regarding $\eta_{m}$ can be extracted from their 
interpretation of the $\eta_{M}$ term.

It is useful to note that we have $\rho_{M} < 0.75$ in our 
model with the choice $A = A_{1 \, 1}$ for either 
neutrino type.  This follows from the relation  
$(\rho_{c} + \varepsilon_{m} \rho_{m}) > 0$, as seen in 
Eqs.~(\ref{eq:3no09}) and (\ref{eq:3no12}).  But 
the mean value in Eq.~(\ref{eq:3no20}) is 
$\rho_{M} = 0.75080$. Accordingly, this implies  
$(\rho_{c} + \varepsilon_{m} \rho_{m}) < 0$, 
although $\rho_{M} < 0.75$ is satisfied within experimental 
uncertainty.%
%%%%% %%%%%  Footnote begin  %%%%% %%%%%
\footnote{In order to account for the old result,
$\rho_{M} = 0.7518 \pm 0.0026$, larger values of 
$|g_{LL}^{S}|$ and $|g_{RL}^{S}|$ have been 
obtained in Ref.~\citen{Fetscher}.}
%%%%% %%%%% Footnote end  %%%%% %%%%%

Thus, we cannot distinguish from this experimental result 
whether the neutrino is of the Dirac or Majorana type.  
Although the $\eta_{m}$ parameter in Eq.~(\ref{eq:3no08}) is 
characteristic of the Majorana neutrino, this parameter 
cannot be used for this distinction, because not only is it multiplied 
by the small coefficient $x_{0}$, but also $\eta_{m}$ 
itself takes a very small value in our model.

However, there is a method that might make it possible to 
distinguish between the two neutrino types.  
This method makes use of the different 
$x$ dependences of the coefficients for $\rho_{c}$ and 
$\rho_{m}$, as seen from Eq.~(\ref{eq:3no08}).  For 
example, suppose we analyze experimental data by using 
Eq.~(\ref{eq:3no08}) with $\rho_{m} \ne 0$ and  obtain 
some $\chi^{2}$ value, say, $\chi^{2}_{m}$ for 
the Majorana neutrino case.  Then, suppose we repeat 
a similar analysis using Eq.~(\ref{eq:3no08}) with 
$\rho_{m} = \eta_{m} = 0$ and thereby determine 
$\chi^{2}_{d}$ for the Dirac neutrino case.  
Indeed, such a $\chi^{2}_{d}$ was already determined by 
the TWIST group.\cite{TwistRho} \ In any case, if $\chi^{2}_{m}$ 
is much smaller than $\chi^{2}_{d}$, we can conclude 
that there is a higher probability that neutrinos are of 
the Majorana type.

As an aside, it is worthwhile noting that $\rho_{c}$ and 
$\rho_{m}$ appear symmetrically.  For example, we have the same 
coefficient for $k_{\pm \, c}$ and $k_{\pm \, m}$ in 
Eq.~(\ref{eq:3no01}), which corresponds to the choice 
$A = A_{0 \, 0}$.  Thus, the choice $A = A_{n \, n}$ is 
not of interest to us, because it does not discriminate 
between the Dirac and Majorana neutrino cases.  On the other 
hand, if we choose $A = A_{0 \, 1}$, 
we have an expression similar to Eq.~(\ref{eq:3no08}) for 
$A = A_{1 \, 0}$.  That is to say, the roles of 
$\rho_{c}$ and $\rho_{m}$ are exchanged.  Of course, 
theoretically, $A_{1 \, 0}$ and $A_{0 \, 1}$ are different, 
but it is not easy to distinguish 
them experimentally from the total decay rate, 
because their deviations from unity 
seem to be small.

We now discuss an important property useful for choosing $A$ 
when we try to compare $\chi^{2}_{m}$ with 
$\chi^{2}_{d}$.  It is desirable to make the difference 
between $\chi^{2}_{m}$ and $\chi^{2}_{d}$ as large as 
possible.  Because this difference depends on the different 
$x$ dependences of terms including the $\rho_{c}$ and 
$\rho_{m}$ parameters, this situation can be realized by 
choosing $A_{n \, \ell}$ with $n \ne \ell$, as seen in the 
derivation of Eq.~(\ref{eq:3no07}).  Let us fix $\ell =0$ for 
simplicity. Then, the $x$ dependence of the term including 
$\rho_{m}$ is $12x(1-x)$, as shown 
in Eq.~(\ref{eq:3no08}), while the $x$ dependence of the term 
including $\rho_{c}$ is listed for various 
$n$ in Table I.  Therefore, we have the  
freedom to minimize the $\chi^{2}$ value by choosing 
$A_{n \, \ell}$ according to the 
pattern of data distribution.  Despite this fact, hereafter, we 
use the choice $A = A_{1 \, 0}$ in this and subsequent 
sections to simplify our description.

\begin{table}[ht]
\caption{The $x$ dependence of the term including 
$\rho_{c}$ for the cases of various $A_{n \, 0}$.}
\label{tab:3no1}
\begin{center}
\begin{tabular}{ccl} \hline \hline
\hspace{2mm} $n$ \hspace{2mm} &
\hspace{2mm} $A_{n \, 0}$ \hspace{2mm} &
Term including $\rho_{c}$ \hspace{3mm} \\ \hline
0   & $A_{0 \, 0}$ & $12x(1-x)$ \\
1   & $A_{1 \, 0}$ & $2x(3-4x)+2x_{0}^{2}$ \\
2   & $A_{2 \, 0}$ & $-4x^{2}+4x_{0}^{2}$ \\
3   & $A_{3 \, 0}$ & $-6x+6x_{0}^{2}$ \\
4   & $A_{4 \, 0}$ & $-4x(3-x)+8x_{0}^{2}$ \\ \hline
\end{tabular}
\end{center}
\end{table}

\subsection{\label{sec:3no2}Anisotropic part of the spectrum: $P(x)$}

Next let us examine the features of $P(x)$ and introduce 
the following quantity for the common factor:
\begin{eqnarray}
B_{n \, \ell} = ( a_{-} + 2 n \, k_{- \, c}
      + 2 \ell \, \varepsilon_{m} k_{- \, m} \, ).
\label{eq:3no21}
\end{eqnarray}

We now show that Eq.~(\ref{eq:3no05}) is a 
special case of Eq.~(\ref{eq:3no02}) with the choice of 
either $B_{3 \, 0}$ or $B_{3 \, 3}$.  First, in the 
case of $B_{3 \, 0}$, we obtain the expression
\begin{eqnarray}
P(x) &=& \xi \sqrt{x^{2} - x_{0}^{2}}
     \Bigl\{ ( -1 + 2 x - r_{0}^{2} )
     + 6 \, \delta_{c}
     \, \left( 3 - 4 x + r_{0}^{2} \right)
    + 12 \, \varepsilon_{m} \,
      \delta_{m} \, (1 - x) \Bigr\}, \nonumber \\
      \label{eq:3no22}
\end{eqnarray}
\noindent
where the parameters are defined as follows:
\begin{eqnarray}
\xi = \left( \frac{B_{3 \, 0}}{A_{1 \, 0}} \right),
     \hspace{3mm} 
\delta_{c} = \left( \frac{k_{- \, c}}{B_{3 \, 0}}
     \right) 
     \hspace{3mm} \mbox{and} \hspace{3mm}
\delta_{m} =
     \left( \frac{k_{- \, m}}{B_{3 \, 0}} \right) .
      \label{eq:3no23}
\end{eqnarray}
It is easy to obtain Eq.~(\ref{eq:3no05}) from 
Eq.~(\ref{eq:3no22}) by introducing the relations
\begin{eqnarray}
6 \, \delta_{c} = \left( 1 - \frac{4}{3} \, \delta_{M} \right)
   \hspace{5mm} \mbox{and} \hspace{5mm} \xi = \xi_{M},
\label{eq:3no26}
\end{eqnarray}
\noindent
because we do not take account 
of the term $\delta_{m}$ appearing only in the Majorana neutrino 
case.

Next, in the case with $B_{3 \, 3}$, we obtain the 
very same form as Eq.~(\ref{eq:3no05}), with 
$\delta_{c}$ in Eqs.~(\ref{eq:3no22}) and 
(\ref{eq:3no26}) replaced by 
$(\delta_{c} + \varepsilon_{m} \delta_{m})$.  Of course, the 
constant $B_{3 \, 0}$ 
in Eq.~(\ref{eq:3no23}) 
should be replaced by $B_{3 \, 3}$.

It is worthwhile noting that $\delta_{c}$ and 
$\delta_{m}$ also appear symmetrically.  Therefore, the 
situation is quite similar to that with the choice of the normalization 
factor $A = A_{n \, \ell}$.  It is possible to obtain  
different $x$ dependences of the terms including 
$\delta_{c}$ and $\delta_{m}$ by choosing a common constant 
$B_{n \, \ell}$ with $n \ne \ell$.  By again fixing $\ell =0$ 
for simplicity, the $x$ dependence of the term including 
$\delta_{c}$ is tabulated for various values of $n$ in Table II.  
However, this property 
is not so effective in this case of $P(x)$, in contrast to the case of 
$N(x)$, because $\delta_{c}$ and $\delta_{m}$ themselves become 
zero or very small, as discussed in the next 
paragraph.

\begin{table}[ht]
\caption{The $x$ dependence of the term including $\delta_{c}$ 
for the cases of various $B_{n \, 0}$.}
\label{tab:3no2}
\begin{center}
\begin{tabular}{ccl} \hline \hline
\hspace{2mm} $n$ \hspace{2mm} &
\hspace{2mm} $B_{n \, 0}$ \hspace{2mm} &
Term including $\delta_{c}$ \hspace{2mm} \\ \hline
0   & $B_{0 \, 0}$ & $12(1-x)$ \\
1   & $B_{1 \, 0}$ & $2(7-8x+r_{0}^{2})$ \\
2   & $B_{2 \, 0}$ & $4(4-5x+r_{0}^{2})$ \\
3   & $B_{3 \, 0}$ & $6(3-4x+r_{0}^{2})$ \\
4   & $B_{4 \, 0}$ & $4(5-7x+2r_{0}^{2})$ \\ \hline
\end{tabular}
\end{center}
\end{table}

If the $SU(2)_{L} \times SU(2)_{R} \times U(1)$ model is 
assumed, then the relation 

\begin{eqnarray}
\kappa = \eta
     \label{eq:3no33}
\end{eqnarray}
is obtained from Eq.~(\ref{eq:Ano20}). 
Therefore, we can conclude that in the Dirac neutrino case, 
we have
\begin{eqnarray}
\delta_{c} = 0 ,
     \label{eq:3no34}
\end{eqnarray}
as seen from Eqs.~(\ref{eq:3no23}) and (\ref{eq:3no14}).  
On the other hand, in the Majorana neutrino case, 
the parameters ($\delta_{c}$ and $\delta_{m}$) include 
some very small contributions coming 
from the differences between small values of the lepton 
mixing matrices, as seen from Eqs.~(\ref{eq:3no17}) and 
(\ref{eq:3no18}). 
Thus in this case, although strictly speaking 
they are non-zero, for practical purpose we can approximate 
them as zero: 
\begin{eqnarray}
\delta_{c} \simeq 0 \hspace{5mm} \mbox{and} 
     \hspace{5mm} \delta_{m} \simeq 0 .
     \label{eq:3no35}
\end{eqnarray}

From these relations, we can conclude within the $SU(2)_{L} \times 
SU(2)_{R} \times U(1)$ model that we have the following
expression:
\begin{eqnarray}
P(x) = \xi \sqrt{x^{2} - x_{0}^{2}}
     ( -1 + 2 x - r_{0}^{2} ) .
      \label{eq:3no36}
\end{eqnarray}
If we set $\xi = 1$, this $P(x)$ itself is that obtained 
from the standard model.  Although this form is common to 
the Dirac and Majorana neutrino cases, it should be noted 
that the definition of $\xi$ differs in the two cases, as seen from 
Eqs.~(\ref{eq:3no23}), (\ref{eq:3no14})~--~(\ref{eq:3no17}), 
namely,
\begin{eqnarray}
\xi & = & \frac {(1 - \lambda^{2})}
     {(1 + \lambda^{2} + 2 \, \eta^{2})}
     \hspace{5mm} \mbox{for the Dirac neutrino case,} 
     \label{eq:3no37} \\
\xi & \simeq & 1
     \hspace{30mm} \mbox{for the Majorana neutrino case.}
     \label{eq:3no38}
\end{eqnarray}
\noindent
Here, the deviation from $\xi = 1$ for the Majorana neutrino case 
is not expressed explicitly, because it cannot be measured 
within the present experimental 
precision.  This definition of $\xi$ is 
independent of the choice of $B_{n \, \ell}$ in this 
$SU(2)_{L} \times SU(2)_{R} \times U(1)$ model.

Note that the Michel parameter $\delta_{M}$ 
should be 
\begin{eqnarray}
     \delta_{M} = 0.75,
     \label{eq:3no39}
\end{eqnarray}
\noindent
within the present experimental precision, as seen 
from Eq.~(\ref{eq:3no26}), because 
$\delta_{c} \simeq 0$ in this model.  The values $\delta_{M} = 0.75$ 
and $\xi = 1$ seem to be consistent with the recent experimental 
results obtained by the TWIST group;\cite{TwistDelta}
\begin{eqnarray}
  & & \delta_{M} = 0.74964 \pm 0.00066 \pm 0.00112,
  \label{eq:3no40} \\
  & & 0.9960 < P_{\mu} \, \xi \le \xi < 1.0040.
     \label{eq:3no41}
\end{eqnarray}

Under the assumption that neutrinos are of the Dirac type, 
this group estimated 
$\lambda_{c} < (80.425/420)^{2} = 3.7 \cdot 10^{-2}$ 
from Eqs.~(\ref{eq:3no41}) and (\ref{eq:3no37}), where 
$\lambda_{c}$ is approximately equal to the ratio of the mass 
of $W_{L}$ to the mass of $W_{R}$ squared [see Eq.~(\ref{eq:Ano19})].

\subsection{Summary for the spectrum}

Let us summarize our expression for the spectrum.  
Because it consists of many terms, we ignore  
some small terms, like $x_{0}$ and $r_{0}^{2}$, 
in order to see the essential features.  Furthermore, 
in order to see its characteristic features, 
we assume the $SU(2)_{L} \times 
SU(2)_{R} \times U(1)$ model; in other words, we 
ignore the $\delta_{c}$ and $\delta_{m}$ parameters 
in Eq.~(\ref{eq:3no22}).  Then, we have the expression
\begin{eqnarray}
D(x, \, \theta) 
  &=& x \,
     \left[ (3 - 2 x) + 2 \, \rho_{c} \, (3 - 4 x)
     + 12 \, \varepsilon_{m} \, \rho_{m} \, (1 - x)
     \pm P_{\mu} \, \xi \, \cos \theta \,( - 1 + 2 x ) \right],
     \nonumber \\
\label{eq:3no42}
\end{eqnarray}
\noindent
where the normalization factor $A = A_{1 \, 0}$ has 
been used.  If another $x$ dependence of the term including 
the $\rho_{c}$ (or $\rho_{m}$) parameter is chosen 
according to Table I, all $A_{1 \, 0}$ 
in this section should be replaced by the 
corresponding $A_{n \, \ell}$.

Finally, we mention the theoretical expression 
for the experiment which 
determined the following quantity:
\begin{eqnarray}
\omega \equiv P_{\mu} \displaystyle \lim_{x \to 1}
     \left[ \frac{P(x)}{N(x)} \right].
\label{eq:3no43}
\end{eqnarray}
The experimental result
\begin{eqnarray}
\omega > 0.99682
\label{eq:3no44}
\end{eqnarray}
was reported by Jodidio et al.\cite{Jodidio}

Here, it should be noted that the real allowed range of $x$ 
is limited by $x_{\mathrm{max}}=(1-r_{jk}^2)$, as shown in 
Eq.~(\ref{eq:2no08}).  Also, both $N(x)$ and $P(x)$ become 
zero at $x_{\mathrm{max}}$, as mentioned in 
Eq.~(\ref{eq:2no17}).  However, since these 
restrictions are only effective in a very tiny range, 
it is understood that this definition of $\omega$ is a  
theoretical result obtained by taking the extrapolation from the 
allowed range of $x$.  In addition, because the radiative 
correction is known to be larger for 
$x > 0.9$, as shown in Ref.~\citen{Arbuzov}, it is assumed that 
the experimental results are adjusted by taking this
radiative correction into account.

Because this $\omega$ is defined by taking the ratio of 
two parts of the $e^{\pm}$ spectrum, it is independent of the choice 
of the normalization factor $A_{n \, \ell}$.  In other 
words, its theoretical expression is obtained from 
Eqs.~(\ref{eq:3no01}) and (\ref{eq:3no02}) as follows:
\begin{eqnarray}
\omega = P_{\mu} \,\, \frac{a_{-}}{a_{+}}
     = P_{\mu} \,\, \frac{\xi_{M} \, \delta_{M}}
     {\rho_{M}},
     \label{eq:3no45}
\end{eqnarray}
\noindent
where the coefficients $a_{\pm}$ are defined in Eqs.~(\ref{eq:3no14}) 
and (\ref{eq:3no16}), 
and the last expression is obtained from the Michel 
parameterization given in Eqs.~(\ref{eq:3no04}) and 
(\ref{eq:3no05}).  For  simplicity, 
the contribution from the electron mass is not included here 
by setting $x_{0}^{2} = r_{0}^{2} = 0$.

For convenience, we now present explicit forms of $\omega$  
in our model.  In the Dirac neutrino case, it is
\begin{eqnarray}
\omega = P_{\mu} 
     \frac{1 - \lambda^{2}}{1 + \lambda^{2}},
     \label{eq:3no46}
\end{eqnarray}
\noindent
while in the Majorana neutrino case, it becomes
\begin{eqnarray}
\omega = P_{\mu} 
     \frac{\left( 1 - \overline{u_{e}}^{\, 2} \right)
     \left( 1 - \overline{u_{\mu}}^{\, 2} \right)
     - \lambda^2 \, \overline{v_{e}}^{\, 2} \,
     \overline{v_{\mu}}^{\, 2}}
     {\left( 1 - \overline{u_{e}}^{\, 2} \right)
     \left( 1 - \overline{u_{\mu}}^{\, 2} \right)
     + \lambda^2 \, \overline{v_{e}}^{\, 2} \,
     \overline{v_{\mu}}^{\, 2}}
     \simeq P_{\mu}.
     \label{eq:3no47}
\end{eqnarray}
\noindent
Here, the last approximation is good in practice, 
because $\overline{u_{\ell}}^{\, 2}$ and 
$\overline{v_{\ell}}^{\, 2}$ seem to be very small, as shown 
in Eq.~(\ref{eq:2no21}).  In other words, no deviation 
from the standard model can be expected again in the Majorana 
neutrino case, within the present experimental precision.
%

%%%%%%%%%%%%%%%%%%%%%%%%%%%%%%%%%%%%%%%%%%%%%%%%%%%%%%%%%%%%%%%
%%%%%%%%%%%%%%%%%          Section 4         %%%%%%%%%%%%%%%%%%
%%%%%%%%%%%%%%%%%%%%%%%%%%%%%%%%%%%%%%%%%%%%%%%%%%%%%%%%%%%%%%%
\section{\label{sec:four}Polarization of $e^{\pm}$}

We now define three components of the spin polarization 
of $e^{\pm}$, $\vec{P_{e}}(x, \, \theta)$ in 
Eq.~(2$\cdot$5).  Its longitudinal component along the momentum 
direction ($\vec{p_e}$) is expressed as 
$P_{L}(x,\theta)$.  In order to separate its transverse components, 
we choose the decay plane defined by this $\vec{p_e}$ and 
the muon polarization vector ($\vec{P}_{\mu}$).  The components 
of the transverse polarization within and perpendicular to this 
decay plane are, respectively, expressed as 
$P_{T1}(x,\theta)$ and $P_{T2}(x,\theta)$.  Mathematically, 
these three components 
are expressed as follows:\cite{Fetscher}
\begin{equation}
  \vec{P}_{e}(x,\theta) 
     = P_L(x,\theta) \hat{p_e}
     + P_{T1}(x,\theta)
       \frac{(\hat{p_e}\times\vec{P}_{\mu})\times\hat{p_e}}
       {|(\hat{p_e}\times\vec{P}_{\mu})\times\hat{p_e}|}
     + P_{T2}(x,\theta)
       \frac{\hat{p_e}\times\vec{P}_{\mu}}
       {|\hat{p_e}\times\vec{P}_{\mu}|}.
  \label{eq:4no01}
\end{equation}

It is convenient to separate the $x$-dependent parts of 
these components from the emission angle of $e^{\pm}$, 
namely, $\cos \theta = (\hat{p_e} \cdot \hat{P}_{\mu})$. 
Therefore, we introduce the quantities
\begin{eqnarray}
P_{L}(x,\theta)
  &=&  \frac{\pm \, Q(x) + P_{\mu} \, \cos\theta \, S(x)}
     {D(x, \, \theta)}, 
     \label{eq:4no02} \\
P_{T1}(x,\theta)
  &=&  \frac{P_{\mu} \, \sin\theta \, R(x)}
     {D(x, \, \theta)},
     \label{eq:4no03} \\
P_{T2}(x,\theta)
  &=&  \frac{P_{\mu} \, \sin\theta \, T(x)}
     {D(x, \, \theta)},
     \label{eq:4no04}
\end{eqnarray}
\noindent
where the denominator $D(x, \, \theta)$ is given 
in Eq.~(\ref{eq:3no42}), if the $SU(2)_{L} \times 
SU(2)_{R} \times U(1)$ model is assumed.

The explicit expressions of these $Q(x)$, $S(x)$, $R(x)$ and 
$T(x)$ are presented in terms of the 
parameters defined in \S3.  Strictly 
speaking, if $A = A_{n \, \ell}$ is chosen in the case 
of $N(x)$, $A_{1 \, 0}$ appearing 
in this section should be replaced by $A_{n \, \ell}$.

\subsection{Longitudinal polarization: $Q(x)$ and $S(x)$}

The isotropic part, $Q(x)$, and anisotropic part, $S(x)$, 
of the longitudinal polarization are, respectively, expressed as 
follows under Conditions (A) and (B) given in \S\ref{sec:two}:
\begin{eqnarray}
Q(x) &=& \xi \sqrt{x^{2} - x_{0}^{2}}
      \left[ (3 - 2 \, x - r_{0}^{2})
     - 6 \, \delta_{c} ( 1 - r_{0}^{2} )
     + 12 \, \varepsilon_{m} \,
      \delta_{m} \, (1 - x) \right],
     \label{eq:4no05} \\
S(x) &=& \left[ ( -x + 2 \, x^{2} - x_{0}^{2} )
      + 2 \, \rho_{c} \, ( 7 \, x - 8 \, x^{2} + x_{0}^{2} )
      \right. \nonumber \\
  & & {} \hspace{13mm} \left. + 12 \, \varepsilon_{m} \,
      \rho_{m} \, x ( 1 - x ) - 2 \, \varepsilon_{m} \,
      \eta_{m} \, x_{0} \, ( 1 - x ) \right].
     \label{eq:4no07}
\end{eqnarray}
\noindent
The parameters ($\xi$, $\delta_{c}$ and $\delta_{m}$) in $Q(x)$ 
are defined in Eq.~(\ref{eq:3no23}) 
for the case of $P(x)$.  In other words, the common factor 
$B_{3 \, 0}$ has been used.%
%%%%% %%%%%  Footnote begin  %%%%% %%%%%%%%%%%%%%%%%%%%%%%%%%%%%%%
\footnote{This function $Q(x)$ is equal to $6 \, F_{IP}(x)$ of 
Ref.~\citen{Fetscher}, where one new parameter,
$\xi_{M}^{\prime}$, was introduced,  because the common factor 
$B_{1 \, 0}$ was used instead of our $B_{3 \, 0}$, and the 
interference term between the $(S \pm P)$ and $T$ interactions 
was included in $\xi_{M}^{\prime}$ in order to obtain a compact 
expression.  Therefore, 
$F_{IP}(x)$ has a different $x$ dependence.}
%%%%% %%%%% Footnote end  %%%%% %%%%%%%%%%%%%%%%%%%%%%%%%%%%%%%%%%%%%
In addition, the parameters ($\rho_{c}$, $\rho_{m}$ and 
$\eta_{m}$) in $S(x)$ are defined in Eq.~(\ref{eq:3no09}) 
for the case of $N(x)$.%
%%%%% %%%%%  Footnote begin  %%%%% %%%%%%%%%%%%%%%%%%%%%%%%%%%%%%
\footnote{This function $S(x)$ is equal to $6 \, F_{AP}(x)$ of 
Ref.~\citen{Fetscher}, where one new parameter,
$\xi_{M}^{\prime\prime}$, was introduced, because the common 
factor $A_{3 \, 0}$ was used instead of our $A_{1 \, 0}$, and 
also the interference term between the $(S \pm P)$ and $T$ 
interactions was included in $\xi_{M}^{\prime\prime}$.  Therefore, 
$F_{AP}(x)$ has a different 
$x$ dependence than our $S(x)$.}
%%%%% %%%%% Footnote end  %%%%% %%%%%%%%%%%%%%%%%%%%%%%%%%%%%%%%%%%%%%
Of course, the corresponding results from the standard model 
are obtained by setting $\xi = 1$ and all other parameters to 
zero.  The original forms of 
$Q(x)$ and $S(x)$ before adopting Conditions (A) 
and (B) are presented in Eqs.~(\ref{eq:Bno12}) and 
(\ref{eq:Bno15}) of Appendix B.

If the $SU(2)_{L} \times SU(2)_{R} \times U(1)$ model 
is assumed, $Q(x)$ is reduced to the following simple expression
corresponding to Eq.~(\ref{eq:3no36}):
\begin{eqnarray}
Q(x) =
 \xi \sqrt{x^{2} - x_{0}^{2}} \,
    (3 - 2 \, x - r_{0}^{2}).
     \label{eq:4no06}
\end{eqnarray}
Thus, the longitudinal polarization is expressed as follows, if 
some small terms, like $x_{0}$, $r_{0}^{2}$, $\delta_{c}$ 
and $\delta_{m}$, are ignored in order to see the 
essential features:
\begin{eqnarray}
P_{L}(x, \, \theta) 
  &=&  \left( \frac{x}{D(x, \, \theta)} \right) 
     \left\{ \pm \xi \, (3 - 2 \, x )
     \right. \nonumber \\
  & & {} \hspace{8mm} \left.
     - P_{\mu} \, \cos \theta \,
     \left[ ( 1 - 2 \, x )
     - 2 \, \rho_{c} \, ( 7 - 8 \, x )
     - 12 \, \varepsilon_{m} \,
      \rho_{m} \, ( 1 - x ) \right]
      \right\}.
     \label{eq:4no08}
\end{eqnarray}
Therefore, the longitudinal polarization of $e^{\pm}$ from an unpolarized 
muon (or $\theta = \pi/2$) is expressed as follows in the 
$SU(2)_{L} \times SU(2)_{R} \times U(1)$ model:
\begin{eqnarray}
P_{L}(x, \, \pi/2) 
  &=& \pm \xi \left[ 1 - \frac{2 \rho_{c} (3 - 4 \, x )}
     {(3 - 2 x) + 2 \rho_{c} (3 - 4 x)} \right]
     \hspace{3mm} \mbox{for the Dirac neutrino},
     \label{eq:4no08a} \\
P_{L}(x, \, \pi/2)
  &=& \pm \xi 
    \hspace{45mm} \mbox{for the Majorana neutrino},
     \label{eq:4no08b}
\end{eqnarray}
where $\xi$ is given in Eqs.~(\ref{eq:3no37}) and 
(\ref{eq:3no38}) for the respective cases, 
and the additional terms due to the parameters $\rho_c$ and $\rho_m$ 
have been ignored for the Majorana neutrino case, because 
of their smallness in comparison with the present experimental precision.

Burkard et al.\cite{Burkard} reported the following 
experimental result for $e^{+}$ from an unpolarized muon 
by assuming $\rho_{c} = 0$ for the Dirac neutrino case:
\begin{eqnarray}
P_{L}(x, \, \pi/2) = 0.998 \pm 0.045.
     \label{eq:4no08c}
\end{eqnarray}
The present average value given in the Particle Data 
Group is $P_{L} = 1.00 \pm 0.04$.\cite{Fetscher}

\subsection{Transverse polarization within the decay plane: 
$R(x)$}

Under Conditions (A) and (B) presented in \S\ref{sec:two}, 
we have
\begin{eqnarray}
R(x)
  = \left[ - ( 1 - 14 \, \rho_{c} -
      12 \, \varepsilon_{m} \, \rho_{m}) \, x_{0} (1 - x )
      -2\,  \varepsilon_{m} \,
      \eta_{m} \, ( x - x_{0}^{2} ) \right].
     \label{eq:4no09}
\end{eqnarray}
The result of the standard model is obtained 
by setting $\rho_{c} = \rho_{m} = \eta_{m} = 0$.  The 
original form before adopting Conditions (A) 
and (B) is presented in Eq.~(\ref{eq:Bno18}).%
%%%%% %%%%%  Footnote begin  %%%%% %%%%%%%%%%%%%%%%%%%%%%%%%%%%%
\footnote{ Our $R(x)$ is equal to $6 \, F_{T1}(x)$ 
of Ref.~\citen{Fetscher}.}
%%%%% %%%%% Footnote end  %%%%% %%%%%%%%%%%%%%%%%%%%%%%%%%%%%%%%%

In the Dirac neutrino case, we cannot expect 
any useful information from this measurement, because 
the first main term obtained from the standard model is 
already proportional to the small $x_{0}$.  
By contrast, the Majorana parameter 
$\eta_{m}$ is accompanied by the coefficient $(x - x_{0}^{2})$, 
which  is significantly larger than the coefficient $x_{0}(1-x)$ 
in $N(x)$ [cf. Eq.~(\ref{eq:3no08})].  However, 
$\eta_{m}$ itself is small, as discussed 
in the next paragraph.  Therefore, it is rather difficult to 
obtain definite information regarding $\eta_{m}$ 
from this measurement.

Let us estimate the order 
of magnitude of the parameter $\eta_{m}$, which includes the 
following combination of lepton mixing matrices: 
\begin{eqnarray}
\lambda \, d_{r} = \lambda \, \mbox{Re}
     \mathrm{\Sigma}_{jk}^{ \, \prime} \,
     ( U_{ej}^{*} V_{ek} V_{\mu j}^{\, *} U_{\mu k} )
     = \lambda \, \mbox{Re}
          ( \overline{w_{e \mu}}^{\,\, *} \,\,
     \overline{w_{e \mu \, h}} ),
     \label{eq:4no10}
\end{eqnarray}
where Eqs.~(\ref{eq:3no09}), (\ref{eq:3no19}) and (\ref{eq:2no23}) 
have been used.

Since it can be naturally assumed 
that there are no contributions from the heavy Majorana 
neutrinos, we are able to express 
$\overline{w_{e \mu}}$ as follows by omitting 
the second component of the neutrino mixing matrices in 
Eqs.~(\ref{eq:Ano14}) and (\ref{eq:Ano15}):
\begin{eqnarray}
\overline{w_{e \mu}}
  = {\mathrm{\Sigma}}_{j}^{ \, \prime} \, U_{ej} \, V_{\mu j}
     = {\mathrm{\Sigma}}_{j}^{ \, \prime}
     \left( U_{E}^\dagger U_{\nu}^{(1)} \right)_{ej} 
     e^{-i\varphi}
    \left( V_{E}^\dagger V_{\nu}^{(1)} \right)_{\mu j}.
    \label{eq:4no13}
\end{eqnarray}
In this expression, the first matrix element, $U_{ej}$, is known 
to be of order unity from neutrino oscillation
experiments,\cite{neutrinomass} \ while, concerning 
the second element $V_{\mu j}$, there 
is no reliable information at present.  But if we assume 
the seesaw mechanism, we must consider $(V_{\nu}^{(1)})_{j^{'} j}$ 
to have a very small value, 
as shown in Eq.~(\ref{eq:Ano07}).  We may get some rough idea of its 
order of magnitude from the neutrinoless 
double beta decay, which gives the following upper bound for 
a similar quantity:\cite{Vogel}\cite{Doi1}
\begin{eqnarray}
\langle\lambda\rangle = \lambda \, \left| {\mathrm{\Sigma}}_{j}^{ \, \prime} \,
     U_{ej} V_{ej} 
     ( \cos \theta_{c}^{'} /
     \cos \theta_{c}) \right|
     < {O}(10^{-6}).
     \label{eq:4no14}
\end{eqnarray}
The primed sum in this case represents a sum that extends 
over only the light neutrinos ($m_{j} < 10$ MeV), so that 
contributions from heavier neutrinos are ignored in comparison 
with the virtual 
Majorana neutrino momentum.  Here, 
$\theta_{c}$ and $\theta_{c}^{'}$ are, respectively, 
the Cabibbo-Kobayashi-Maskawa mixing angles for the 
left-handed and right-handed $d$ and $s$ quarks.  The 
order of magnitude of 
$\lambda \, \overline{w_{e \mu}}$ in Eq.(\ref{eq:4no10}) 
seems to be less 
than $10^{-6}$, although the suffix $\mu$ in Eq.~(\ref{eq:4no13}) 
is replaced by the suffix $e$ in Eq.~(\ref{eq:4no14}), and 
there are some quantities related to the quark sector.  Thus 
the order of magnitude of $\eta_{m}$ seems to be much 
smaller than $10^{-6}$, because of the additional factor of 
$\overline{w_{e \mu \, h}}$ in Eq.~(\ref{eq:4no10}).

Recently, Danneberg et al.\cite{TransPol} reported 
the following energy averaged value for $e^{+}$ in the direction 
$\theta = \pi/2$:
\begin{eqnarray}
  \langle P_{T1}(x, \theta = \pi/2)\rangle \, = \,
     (6.3 \pm 7.7 \pm 3.4) \cdot 10^{-3}.
     \label{eq:4no14a}
\end{eqnarray}
This experimental result is of the same order of magnitude 
as the prediction of the standard model,
\begin{eqnarray}
  P_{T1}(x, \theta = \pi/2) = - P_{\mu} \, 
  \frac{x_{0} (1 - x)}{x (3 - 2 x)},
     \label{eq:4no14b}
\end{eqnarray}
where $x_{0}$ is defined in Eq.~(\ref{eq:2no09}).

\subsection{Transverse polarization perpendicular to the 
decay plane: $T(x)$}

A non-zero value of $T(x)$ implies 
the existence of a non-zero Majorana \textit{CP} violation 
phase in our model. It is expressed as  
\begin{eqnarray}
T(x) = 2 \, \varepsilon_{m}
    \sqrt{x^{2} - x_{0}^{2}} \, \eta_{m \, i}
        \left( 1 - r_{0}^{2} \right),
     \label{eq:4no15}
\end{eqnarray}
where the parameter $\eta_{m \, i}$ is defined as follows, 
using Eqs.~(\ref{eq:Bno21}) and (\ref{eq:Bno65}):
\begin{eqnarray}
\eta_{m \, i} = \varepsilon_{m} \, 
     \left( \frac{\lambda \, d_{i}}{A_{1 \, 0}} \right)
     = \varepsilon_{m} \, 
     \left( \frac{\lambda}{A_{1 \, 0}} \right)
     \mbox{Im} ( \overline{w_{e \mu}}^{\,\, *} \,\,
     \overline{w_{e \mu \, h}} ).
     \label{eq:4no16}
\end{eqnarray}
\noindent
There is no corresponding term in either the standard model 
or our model for the Dirac neutrino.%
%%%%% %%%%%  Footnote begin  %%%%% %%%%%%%%%%%%%%%%%%%%%%%%%%
\footnote{Our $T(x)$ is equal to $6 \, F_{T2}(x)$ 
of Ref.~\citen{Fetscher}.  Their $F_{T2}(x)$ includes 
the CP violating term 
even in the massless Dirac neutrino case, because it comes 
from the interference between the $(V \pm A)$ and 
$(S \pm P)$ ( or $T$ ) interactions.}
%%%%% %%%%% Footnote end  %%%%% %%%%%%%%%%%%%%%%%%%%%%%%%%%%
The parameter $\eta_{m \, i}$ is obtained by taking the 
imaginary part instead of the real part in 
Eq.~(\ref{eq:4no10}).  Therefore, $\eta_{m \, i}$ is
proportional to the sin term of the \textit{CP} 
violating phases appearing in the lepton 
mixing matrices.  As we can imagine from Eq.~(\ref{eq:4no14}), 
we cannot expect to measure $\eta_{m \, i}$ in practice, because 
its value seems to be too small, as predicted by our model 
(for details, see \S3 of Ref.~\citen{Doi2}).

Recently, Danneberg et al.\cite{TransPol} reported 
the following energy averaged value for $e^{+}$ in the direction 
$\theta = \pi/2$:
\begin{eqnarray}
  \langle P_{T2}(x, \theta = \pi/2)\rangle \, = \, 
     (-3.7 \pm 7.7 \pm 3.4) \cdot 10^{-3}.
     \label{eq:4no20}
\end{eqnarray}
A smaller value of $\langle P_{T2}(x, \theta = \pi/2)\rangle$ is 
expected if the neutrino is of the 
Majorana type and the $CP$-violating phase exists.
%

%%%%%%%%%%%%%%%%%%%%%%%%%%%%%%%%%%%%%%%%%%%%%%%%%%%%%%%%%%%%%%%
%%%%%%%%%%%%%%%%%          Section 5         %%%%%%%%%%%%%%%%%%
%%%%%%%%%%%%%%%%%%%%%%%%%%%%%%%%%%%%%%%%%%%%%%%%%%%%%%%%%%%%%%%
\section{\label{sec:five}Concluding remarks}

It was shown in \S\ref{sec:three} 
that the Michel parameterization,\cite{Fetscher} which 
has been used by an experimental group, is a 
special case of the more general form to investigate the 
deviation from the standard model.  We propose a new 
parameterization that directly represents deviations from the 
standard model.  In general, there is the freedom to choose 
the normalization factor $A_{n \, \ell}$ in 
Eq.~(\ref{eq:3no06}).  Then, we can find the most effective 
$x$ dependence of the term including that parameter, whose value 
is determined by analyzing experimental data.  The 
$SU(2)_{L} \times SU(2)_{R} \times U(1)$ model has been 
used to elucidate this feature in the simplified expression given 
in Eq.~(\ref{eq:3no42}).  Concerning the spectrum, it has been 
confirmed that there is no significant deviation from the 
standard model, except that the mean value of $\rho_{M}$ 
determined experimentally is slightly larger than 0.75, as shown 
in Eq.~(\ref{eq:3no20}), although we have $\rho_{M} < 0.75$ in 
our model.

Also, it was shown in \S\ref{sec:four} that 
the polarization of $e^{\pm}$ can be expressed in terms of 
the same parameters introduced to analyze 
the spectrum, in contrast to Ref.~\citen{Fetscher}, in which
several new parameters are introduced.  The predictions 
for the polarization obtained from 
the standard model are also consistent with the recent 
experimental results.

It is an important problem to investigate whether 
the neutrino is of the Dirac or Majorana type, as mentioned in 
\S1.  In normal muon decay, there 
are three theoretically possible subjects for this 
purpose within the framework of gauge theory, 
as we now discuss.

The first subject is to measure the transverse polarization of 
$e^{\pm}$ perpendicular to the decay plane, namely $T(x)$ in 
Eq.~(\ref{eq:4no15}).  The reason for this is because this polarization 
exists in neither the standard model nor the massive Dirac neutrino 
case.  However, the theoretical estimate of $T(x)$ 
is very small, as explained below Eq.~(\ref{eq:4no16}).  We cannot 
expect to obtain any useful information from 
this measurement within the present experimental 
precision.

The second subject is the transverse polarization 
in the decay plane, $R(x)$ appearing in Eq.~(\ref{eq:4no09}).  In this 
case, the term associated with $\eta_{m}$ characteristic of the 
Majorana neutrino has the larger $x$ dependence, but this 
$\eta_{m}$ itself is also too small, as mentioned below 
Eq.~(\ref{eq:4no10}).  It seems difficult to derive 
any definite conclusion from this measurement.

The remaining possibility is to take advantage of the 
different $x$ dependences of the terms including the 
parameters $\rho_{c}$ and $\rho_{m}$ in the energy spectrum 
Eq.~(\ref{eq:3no42}) by comparing the 
$\chi^{2}$ values for the Dirac-type neutrino with those 
for the Majorana-type neutrino, as mentioned below 
Eq.~(\ref{eq:3no20}).  This may provide a test 
to determine the type of neutrino, although it is indirect.

Finally, let us summarize the general features. In the 
Dirac neutrino case, there is no important effect due 
to the lepton mixing matrices under Condition (B) in 
Eq.~(\ref{eq:2no19}).  In other words, we can 
use the theoretical expressions obtained by assuming 
massless neutrinos.  In addition, we point 
out that it is useful to choose a different $n$ for the 
normalization factor $A_{n \, 0}$ in Eq.~(3$\cdot$6) in order 
to minimize $\chi^{2}$.  Thus, 
we can find some constraints on the coupling constants 
($\lambda$, $\eta$ and $\kappa$) in principle by combining 
other information from various decay processes.  
We also note that deviations from the standard 
model become smaller as these coupling constants 
become smaller.

In the Majorana neutrino case, it is very difficult to find 
any deviation from the standard model under Condition (B) given 
in Eqs.~(\ref{eq:2no21}) and (\ref{eq:2no23}).  This 
is because all parameters 
include the small components of the lepton mixing
matrices.  This feature is independent of the values of 
$\lambda$, $\eta$ and $\kappa$.

\appendix
%%%%%%%%%%%%%%%%%%%%%%%%%%%%%%%%%%%%%%%%%%%%%%%%%%%%%%%%%%%%%%%
%%%%%%%%%%%%%%%%%          Appendix A        %%%%%%%%%%%%%%%%%%
%%%%%%%%%%%%%%%%%%%%%%%%%%%%%%%%%%%%%%%%%%%%%%%%%%%%%%%%%%%%%%%
\section{Summary of Various Mixing Matrices}\label{sc:AT1}

For the purpose of making this paper self-contained, 
here we summarize the theoretical foundations of this work, 
even though they have been discussed many times in the literature 
already.  Many theoretical gauge models beyond the standard 
model have been proposed to analyze normal muon 
decay~\cite{Doi2again}.  Among them, let us consider a model 
that consists of $V-A$ and $V+A$ currents. 
\par
The mass term of leptons in the Lagrangian with $n$ left-handed and 
$n$ right-handed lepton doublets is generally defined by 
\begin{equation}
{\cal{L}}_M = 
  -\overline{{\cal{E}}_R}M_{E} {\cal{E}}_L
  -\frac{1}{2}\left(\overline{(\nu_{L})^{c}},
  \overline{(\nu_{R}^{\,\prime})}\right){\cal{M}} 
  \left(\begin{array}{cc} \nu_{L} \\ (\nu_{R}^{\,\prime})^{c} 
  \end{array} \right)
  +\mbox{H.c.} ,
\end{equation}
where $\cal{E}$, $\nu_{L}$, and $\nu_{R}^{\,\prime}$ are, 
respectively, the weak eigenstates of the charged leptons and 
left-handed and right-handed neutrinos.  Explicitly, we write 
$\overline{{\cal{E}}_R} = (\overline{e^{\prime}_R}, \overline{\mu^{\prime}_R},
\cdots)$, ${\cal{E}}_L^T = (e^{\prime}_L,\mu^{\prime}_L,\cdots)$, 
$\nu_L^T=(\nu_{eL},\nu_{\mu L},\cdots)$, 
and  $\nu_{R}^{\,\prime \,T} = (\nu_{eR}^{\,\prime},
\nu_{\mu R}^{\,\prime},\cdots)$. 
Note also that $(\nu_{\ell L(R)})^{c} = 
C\, \overline{\nu_{\ell L(R)}}^{\, T}$, where 
$C$ is the charge conjugation operator. 
Here $M_{E}$ is the $n \times n$ mass matrix for charged 
leptons and ${\cal{M}}$ is the $2n \times 2n$ neutrino mass 
matrix defined by 
\begin{equation}
{\cal{M}}=
\left(
        \begin{array}{cc}
        M_L & M_D^T \\
        M_D & M_R 
        \end{array}
\right),
\end{equation}
where $M_D$, $M_L$ and $M_R$ are, respectively, the 
Dirac-type and left-handed and right-handed Majorana type 
$n \times n$ mass matrices for neutrinos. Here, the 
identity $\overline{\nu_{R}^{\,\prime}} M_{D} \nu_{L} = 
\overline{(\nu_{L})^{c}} M_{D}^{T} (\nu_{R}^{\,\prime})^{c}$ 
has been used. 

Let us first examine the case in which the Majorana-type mass 
terms exist.  Since $M_L$ and $M_R$ are symmetric matrices,~\cite{Doi1} 
${\cal{M}}$ is also symmetric and can be diagonalized 
by some orthogonal matrix in 
principle to determine the neutrino 
masses.~\cite{Bilenky}  However, we use 
the $2n \times 2n$ unitary matrix $\cal{U}_{\nu}$ 
in order to obtain positive values for 
masses:\cite{Doi1, Schechter}
\begin{equation}
{\cal{U}}_{\nu}^T{\cal{M}}{\cal{U}}_{\nu}  ={\cal{D}}_{\nu}.
  \label{eq:Ano03}
\end{equation}
Here, ${\cal{D}}_{\nu}$ is a diagonal matrix whose $2n$ 
elements represent the masses of the Majorana-type 
neutrinos. Therefore, the weak eigenstates 
of neutrinos are expressed as superpositions of 
the mass eigenstate Majorana neutrinos $N_{j}$ as follows:
\begin{equation}
\left(
  \begin{array}{cc} \nu_{L} \\ (\nu^{\, \prime}_{R})^{c}
  \end{array} \right) 
= {\cal{U}}_{\nu}N_L
= \left(
  \begin{array}{c} U_{\nu}  \\ V_{\nu}^*
  \end{array} \right)N_L
= \left(\begin{array}{cc} U_{\nu}^{(1)} & U_{\nu}^{(2)} \\
  V_{\nu}^{(1) *} & V_{\nu}^{(2) *} \end{array} \right)
  \left(\begin{array}{c} N_{\mathrm{I}L}  \\ 
  N_{\mathrm{II}L}  \end{array} \right).
  \label{eq:Ano04}
\end{equation}
Thus, we have the $2n$ mass eigenstate Majorana neutrinos, 
$(N_{\mathrm{I}})^T=(N_{1}, N_{2}, \cdots, N_{n})$ 
and $(N_{\mathrm{II}})^T=(N_{n+1}, N_{n+2}, \cdots, 
N_{2n})$.  Here, the $n\times 2n$ neutrino mixing matrices 
$U_{\nu}$ and $V_{\nu}$, which are expressed as
$U_{\nu}=( U_{\nu}^{(1)} , U_{\nu}^{(2)})$ and 
$V_{\nu}=( V_{\nu}^{(1)} , V_{\nu}^{(2)})$,   
are introduced as
%when the weak eigenstate of neutrino $\nu_{\ell L(R)}$ is expressed by   
%the superposition of the mass eigenstate neutrinos $N_j$ with mass $m_j$ 
\begin{equation}
\nu_{\ell L}=\sum_{j=1}^{2n}(U_{\nu})_{\ell j}N_{jL}
     \hspace{5mm} \mbox{and} \hspace{5mm} 
\nu_{\ell R}^{\, \prime} = \sum_{j=1}^{2n}(V_{\nu})_{\ell j}
     N_{jR}
     \label{eq:Ano05}
\end{equation}
for the case of the $n$ generations.~\cite{Doi1}

In this scenario, the small masses of the left-handed Majorana-type 
neutrinos $(N_{\mathrm{I}})$ are naturally explained 
by the seesaw mechanism under the assumption that 
the right-handed Majorana neutrinos $(N_{\mathrm{II}})$ 
have large masses.  Thus, the elements of both $U_{\nu}^{(1)}$ 
and $V_{\nu}^{(2)}$ are of order one, while $U_{\nu}^{(2)}$ 
and $V_{\nu}^{(1)}$ are of order $m_{\nu D}/m_{\nu R}$, 
which seems to be very small. 
Here the quantities $m_{\nu D}$ and $m_{\nu R}$ 
represent the orders of the matrices $M_D$ and $M_R$, respectively:
\begin{eqnarray}
U_{\nu}^{(1)} & =& {O}(1),\quad  \quad  \quad  \quad
  \quad U_{\nu}^{(2)} = {O}(m_{\nu D}/m_{\nu R}),\nonumber\\
V_{\nu}^{(1)} & =& {O}(m_{\nu D}/m_{\nu R}), 
\quad V_{\nu}^{(2)} = {O(1)}.
     \label{eq:Ano07}
\end{eqnarray}
From the unitarity condition for $\cal{U}_{\nu}$, the 
matrices $U_{\nu}^{(1)}$, $U_{\nu}^{(2)}$, $V_{\nu}^{(1)}$ 
and $V_{\nu}^{(2)}$ should satisfy the relations
\begin{eqnarray}
U_{\nu}^{(1)}U_{\nu}^{(1)\dagger} +
     U_{\nu}^{(2)}U_{\nu}^{(2)\dagger} & =& 1,
     \label{eq:Ano08}\\
V_{\nu}^{(1)}V_{\nu}^{(1)\dagger} +
     V_{\nu}^{(2)}V_{\nu}^{(2)\dagger} & =& 1,
     \label{eq:Ano09}\\
U_{\nu}^{(1)}V_{\nu}^{(1)T } +
     U_{\nu}^{(2)}V_{\nu}^{(2)T} & =& 0,
     \quad \mbox{etc.}
\end{eqnarray}
Note that $U_{\nu}^{(1)}$ and $V_{\nu}^{(1)}$ themselves are 
not unitary.

We now note that the reason for the smallness of the quantities 
$\overline{u_{\ell}}^{\, 2}$, 
$\overline{v_{\ell}}^{\, 2}$,  
$\overline{w_{e \mu}}$ 
and $\overline{w_{e \mu \, h}}$ introduced in 
Eqs.~(\ref{eq:2no21}) and (\ref{eq:2no23}) is that  
they include, respectively, the elements of the neutrino 
mixing matrix products 
$(U_{\nu}^{(2)}U_{\nu}^{(2)\dagger})_{\ell \ell}$, 
$(V_{\nu}^{(1)}V_{\nu}^{(1)\dagger})_{\ell \ell}$,
$(U_{\nu}^{(1)}V_{\nu}^{(1)T})_{e \mu}$ and 
$(U_{\nu}^{(1)}V_{\nu}^{(1)T})_{\mu e}$.
In fact, the elements like $U_{\nu}^{(2)}$ and $V_{\nu}^{(1)}$ in them 
seem to be of order $m_{\nu D}/m_{\nu R}$, as shown in 
Eq.~(\ref{eq:Ano07}).

Next, let us consider the other simplified scenario, in 
which $M_{L}=M_{R}=0$. Namely, there exists only the Dirac-type 
neutrino mass matrix $M_D$.  Because there is no theoretical 
restriction on $M_D$ to be symmetric, it can be diagonalized 
by two $n \times n$ unitary matrices $U_{\nu}$ and $V_{\nu}$ as
\begin{equation}
     V_{\nu}^\dagger M_{D} U_{\nu} = {D}_{\nu}^{\,\prime}.
     \label{eq:Ano11}
\end{equation}
Here, ${D}_{\nu}^{\,\prime}$ is a diagonal matrix whose 
$n$ elements represent the masses of the Dirac-type neutrinos. 
Therefore, the weak eigenstates 
of neutrinos in this scenario are expressed as superpositions of 
$n$ mass eigenstate Dirac-type neutrinos $N_{j}$, as shown 
in Eq.~(\ref{eq:Ano05}), although the upper limit of the sum over $j$ 
is restricted to $n$.

In order to avoid the complication of needing to treat the Dirac 
and Majorana neutrino cases separately, 
we use the following convention in this paper:  In the Dirac neutrino 
case, only $n$ mass eigenstate neutrinos $N_{\mathrm{I}}$ exist, 
while $N_{\mathrm{II}}$ do not exist. Thus we can set 
$U_{\nu}^{(2)}=0$ and $V_{\nu}^{(2)}=0$. Therefore, 
$U_{\nu}^{(1)}$ and $V_{\nu}^{(1)}$ are both unitary matrices, and 
they are of order one.  This situation can be 
expressed simply as follows:
\begin{eqnarray}
U_{\nu}^{(1)} & =& {O}(1), \quad U_{\nu}^{(2)} =0,\nonumber\\
V_{\nu}^{(1)} & =& {O}(1), \quad V_{\nu}^{(2)}=0.
     \label{eq:Ano12}
\end{eqnarray}

We have considered the simplified case of introducing the Dirac 
neutrino fields. However, there is another possible scenario. 
In that scenario, 
one Dirac neutrino field can be expressed as a superposition 
of two Majorana neutrino fields with equal masses\cite{Doi4}. 
In this case,  
one Majorana neutrino field is chosen from $N_{\mathrm{I}}$ 
and the other field from $N_{\mathrm{II}}$; that is, we have 
$m_{j+k} = m_{j}$ for $1 \leq j(k) \leq n$. 

As a typical example of gauge theory with
left- and right-handed weak gauge bosons, we consider the 
$SU(2)_L \times SU(2)_R \times U(1)$ model~\cite{Doi1, Herczeg}. In 
this framework, the charged weak current interaction is written in 
the bases of 
mass eigenstates of the charged leptons $E_{\ell}$ and the 
neutrino $N_j$ after diagonalizing the charged lepton and 
neutrino mass matrices:
\begin{eqnarray}
{\cal{L}_{CC}}=& -&\frac{g_L}{2\sqrt{2}}\sum_{\ell,j}
     \overline{E_{\ell}}\gamma_\alpha(1-\gamma_5)
     (U_{E}^\dagger U_{\nu})_{\ell j}N_{jL}W_{L}^{\, \alpha} \nonumber \\
& -&\frac{g_R}{2\sqrt{2}}\sum_{\ell,j}
     \overline{E}_{\ell}\gamma_\alpha(1+\gamma_5)
     (V_{E}^\dagger V_{\nu})_{\ell j}N_{jR}W_{R}^{\, \alpha}
     + \mbox{H.c.}
     \label{eq:Ano13}
\end{eqnarray}
Here, $g_L$ and $g_R$ are the real gauge coupling constants for 
the left- and right-handed weak gauge bosons, $W_L$ and $W_R$, respectively,  
and $U_{E}$ and $V_{E}$ are unitary matrices that diagonalize 
the charged lepton mass matrix $M_{E}$ as 
$V_{E}^\dagger M_{E}U_{E}={D}_{E}$, in analogy to Eq.~(\ref{eq:Ano11}). 

The weak gauge bosons $W_L$ and $W_R$ are expressed in terms 
of the mass eigenstate gauge bosons $W_1$ and $W_2$ as
\begin{eqnarray}
W_L & =& W_1 \cos\zeta + e^{i\varphi} W_2\sin\zeta,
     \label{eq:Ano16} \\
W_R & =& -e^{-i\varphi} W_1 \sin\zeta + W_2\cos\zeta,
     \label{eq:Ano17}
\end{eqnarray}
where $\zeta$ is a $W_L-W_R$ mixing angle and 
$\varphi$ is a $CP$-violating phase. 
The phase factor $e^{-i\varphi}$ comes from 
the VEV of the $(\frac{1}{2},\frac{1}{2},1)$ Higgs field 
through the diagonalization of the mass matrix of the charged weak 
gauge bosons. 

In obtaining the effective current-current interaction,
the left- and right-handed 
Maki-Nakagawa-Sakata (MNS) lepton mixing matrices $U$ and $V$ are 
defined as 
\begin{eqnarray}
U &= & U_{E}^\dagger U_{\nu}=
     \left( U_{E}^\dagger U_{\nu}^{(1)} ,
     U_{E}^\dagger U_{\nu}^{(2)}\right),
     \label{eq:Ano14} \\
V &= & e^{-i\varphi} V_{E}^\dagger V_{\nu}=
     e^{-i\varphi}\left( V_{E}^\dagger V_{\nu}^{(1)} ,
     V_{E}^\dagger V_{\nu}^{(2)}\right).
     \label{eq:Ano15} 
\end{eqnarray}
It should be noted that, 
although the lepton mixing matrices $U$ and $V$ themselves are 
unitary in the Dirac neutrino case,  
the $2n \times 2n$ lepton mixing matrix 
${\small{\Big(
        \begin{array}{c}
        U \\
        V^{*} 
        \end{array}
\Big)}}$ 
is a unitary matrix in the Majorana neutrino case. 
 
There exists some freedom in the treatment of the phase factor 
$e^{-i\varphi}$ mentioned above.  In our treatment, 
this phase factor is absorbed into $V$ of Eq.~(\ref{eq:Ano15}) 
in order to make the coupling constants $\eta$ and $\kappa$ real 
in the effective Hamiltonian of Eq.~(\ref{Hamiltonian}). 
Contrastingly, Herczeg~\cite{Herczeg} includes it in the 
definitions of $\kappa$ and $\eta$.  In our convention, 
all of the constants $G_F$, $\lambda$, $\eta$ and $\kappa$ 
are real:
\begin{eqnarray}
\frac{G_F}{\sqrt{2}} & =& \frac{1}{8}
     \left(\frac{g_{L}\cos \zeta}{M_1}\right)^2
     \left(1+\lambda_c \tan^2 \zeta \right),
     \label{eq:Ano18} \\
\lambda & =& \left( \frac{g_{R}}{g_{L}} \right)^2 
     \frac{\lambda_c +\tan^2 \zeta      }{1+\lambda_c \tan^2 \zeta},
     \label{eq:Ano19} \\
\kappa & =& \eta  =-\left( \frac{g_{R}}{g_{L}} \right)
     \frac{\left(1-\lambda_c \right)\tan \zeta}
     {1+\lambda_c \tan^2 \zeta},
     \label{eq:Ano20}
\end{eqnarray}
where $\lambda_c = (M_{1}/M_{2})^2$, $M_{i}$ being the mass of $W_{i}$.
Note that we have $\lambda \sim \kappa^{2} = \eta^{2}$ for 
$\lambda_c \ll \tan^{2}\zeta < 1$.
%%

%%%%%%%%%%%%%%%%%%%%%%%%%%%%%%%%%%%%%%%%%%%%%%%%%%%%%%%%%%%%%%%
%%%%%%%%%%%%%%%%%         Appendix B         %%%%%%%%%%%%%%%%%%
%%%%%%%%%%%%%%%%%%%%%%%%%%%%%%%%%%%%%%%%%%%%%%%%%%%%%%%%%%%%%%%
\section{Definitions of Various Coefficients}\label{sc:AT2}

Our model given by Eqs.~(\ref{Hamiltonian}) and (\ref{eq:2no03}) differs 
from the standard model because it takes account of the finite neutrino 
mass and the $V+A$ current. 
The complete decay formulas obtained 
by keeping both the neutrino masses ($m_j$ and $m_k$) 
and the $V+A$ parameters ($\lambda$ and $\kappa=\eta$), 
are given in Ref.~\citen{Doi3} 

Terms proportional to the neutrino masses can be omitted in practice, 
because the neutrino masses are very small in comparison with the energy scale $(W)$: 
\begin{equation}
\left(\frac{m_{\nu}}{W}\right)= 1.89 \cdot 10^{-7},
\label{eq:Bno01}
\end{equation}
where $m_{\nu}$ represents $m_j$ and $m_k$, 
and is taken to be 10 eV in order to get a rough idea for the magnitude. 

Then, we have the following results for various components 
of the decay rate:%
%%%%%%%%%%%%%%%%%%%%%  footnote begin  %%%%%%%%%%%%%%%%%
\footnote{The relations between the definitions 
in this paper and in Ref.\,\citen{Doi2} are as follows: 
$A \, N(x)=(x/W)N(e)$,\ 
$A \, P(x)=-(x p_{e}/WE) P(e) $,\ 
$A \, Q(x)=(x \, p_{e}/WE)Q(e)$,\ 
$A \, S(x)=-(x/W)S(e)$,\ 
$A \, R(x)=-(x/W) R(e)$,\ 
$A \, T(x)=-(x p_{e}/WE)T(e)$.}
%%%%%%%%%%%%%%%%%%%  footnote end %%%%%%%%%%%%%%%%%%%%%%
\begin{eqnarray}
N(x) 
&=& \left(\frac{1}{A}\right)\Bigl[a_{+} (3 x - 2 x^{2} - x_{0}^{2})
+ (b_{+} - a_{+}) \, (2 x - x^2 - x_{0}^{2})
\Bigr. \nonumber \\
& & {} \hspace{15mm} 
+ 12 \, k_{+}
\, x \, (1 - x) + 6 \, \varepsilon_{m} \, \lambda \, d_{r} \, 
x_{0} \, (1-x)\Bigl], 
\label{eq:Bno06}\\
P(x)
&=& \left(\frac{1}{A}\right)\sqrt{x^{2} - x_{0}^{2}}
\Bigl[ a_{-} (-1 + 2 \, x - r_{0}^{2})
+ ( b_{-} - a_{-} ) ( x - r_{0}^{2} )
\Bigr. \nonumber \\
& & {} \hspace{35mm} \Bigl.
+ 12 \, k_{-} \, (1 - x) \Bigr],
\label{eq:Bno09}\\
Q(x)
&=& \left(\frac{1}{A}\right)\sqrt{x^{2} - x_{0}^{2}}
\Bigl[ a_{-} (3 - 2 \, x - r_{0}^{2})
+ ( b_{-} - a_{-} ) (2 - x - r_{0}^{2} )
\Bigr. \nonumber \\
& & {} \hspace{35mm} \Bigl.
+ 12 k_{-}(1 - x) \Bigr],
\label{eq:Bno12}\\
S(x)
&=& \left(\frac{1}{A}\right)\Bigl[a_{+} \, ( -x + 2 \, x^{2} - x_{0}^{2}) )
+ ( b_{+} -  a_{+} ) \, ( x^{2} - x_{0}^{2})
\nonumber \\
& & {} \hspace{23mm}
- 12 \, k_{+} \, (-x + x^{2})- 2 \, \varepsilon_{m} \,
\lambda \, d_{r} \, x_{0} \, ( 1 - x )\Bigl],
\label{eq:Bno15}\\
R(x)
&=& \left(\frac{1}{A}\right)
\Bigl[- a_{+} \, x_{0} (1 - x )
+ 12 \, k_{+} \, x_{0} (1 - x )
- 2 \, \varepsilon_{m} \,
\lambda \, d_{r} \, ( x - x_{0}^{2} )\Bigl], 
\label{eq:Bno18}\\
T(x)
&=& \left(\frac{1}{A}\right)2 \, \varepsilon_{m}
\sqrt{x^{2} - x_{0}^{2}}
\, \lambda \, d_{i}
\left( 1 - r_{0}^{2} \right).
\label{eq:Bno21}
\end{eqnarray}
Here, we keep all terms with respect to 
$\lambda,\kappa$ and $\eta$, 
whereas only their first order terms are kept in Ref.~\citen{Doi2}. 
The coefficients in these results are defined as follows:
\begin{eqnarray}
a_{\pm} &=& ( a \pm \lambda^2 \, \hat{a} ),\hspace{25mm}
b_{\pm} = ( b \pm \lambda^2 \, \hat{b} ),
\label{eq:Bno34} \\
k_{\pm} &=& ( \, k_{\pm \, c} + \varepsilon_{m} k_{\pm \, m} \, ),
\label{eq:Bno36} 
\end{eqnarray}
where%
%%%%%%%%%%%%%%%%%%%%%%  footnote begin  %%%%%%%%%%%%%%%%%
\footnote{Here we have introduced the new quantities $k_{\pm},d_{r}$ and 
$d_{i}$ in place of $x_{\pm}, h$ and $h^{I}$ used in 
Ref.~\citen{Doi3}.  The relations between them are 
$\kappa^{2} x_{\pm}=6(k_{\pm c}+\varepsilon_{m} k_{\pm m}), 
h=\varepsilon_{m} d_{r}$ and $h^{I}=\varepsilon_{m} d_{i}$.} 
%%%%%%%%%%%%%%%%%%%%  footnote end  %%%%%%%%%%%%%%%%%%%%
\begin{eqnarray}
k_{\pm \, c} = \frac{1}{2}
\left( \kappa^2 \, c \pm \eta^{2} \, \hat{c} \right),\hspace{10mm}
k_{\pm \, m} = \frac{1}{2}
\left( \kappa^2 \, d \pm \eta^{2} \, \hat{d} \right).
\label{eq:Bno40}
\end{eqnarray}

All of these coefficients are classified into two groups. 
One group consists of $a,b,c,\hat{a},\hat{b}$ and $\hat{c}$, which 
are common to the Dirac and Majorana neutrino cases. The 
definitions of $a,b,\hat{a}$ and $\hat{b}$ are given in 
Eqs. (\ref{eq:2no13a}) and (\ref{eq:2no13b}), while 
$c$ and $\hat{c}$ are given by
\begin{align}
&c = \sum_{jk}{'} \frac{1}{3} F_{j \, k}^{1/2}
( 2F_{j \, k} + G_{j \, k} ) \,
|U_{ej}|^{2} |V_{\mu k}|^{2},\hspace{-25mm}&
\label{eq:Bno25} \\
&\hat{c} = \sum_{jk}{'} \frac{1}{3} F_{j \, k}^{1/2}
( 2F_{j \, k} + G_{j \, k} ) \,
|V_{ej}|^{2} |U_{\mu k}|^{2}.\hspace{-25mm}&
\label{eq:Bno26}
\end{align}
The other group is only for the Majorana neutrino case:
\begin{eqnarray}
d &=& \sum_{jk}{'} \frac{1}{3} F_{j \, k}^{1/2}
( 2F_{j \, k} + G_{j \, k} ) \,
\mbox{Re} ( U_{ej}^{*} U_{ek} V_{\mu j}^{*} V_{\mu k} ),
\label{eq:Bno28} \\
\hat{d} &=& \sum_{jk}{'} \frac{1}{3} F_{j \, k}^{1/2}
( 2F_{j \, k} + G_{j \, k} ) \,
\mbox{Re} (V_{ej}^{*} V_{ek} U_{\mu j}^{*} U_{\mu k} ),
\label{eq:Bno29} \\
d_{r} &=& \sum_{jk}{'} F_{j \, k}^{1/2} G_{j \, k} \,
\mbox{Re} ( U_{ej}^{*} V_{ek} V_{\mu j}^{\, *} U_{\mu k} ),
\label{eq:Bno30} \\
d_{i} &=& \sum_{jk}{'} F_{j \, k}^{1/2} G_{j \, k} \,
\mbox{Im} ( U_{ej}^{*} V_{ek} V_{\mu j}^{\, *} U_{\mu k} ).
\label{eq:Bno31}
\end{eqnarray}

Under Condition (A) in Eq.~(\ref{eq:2no18}), we have 
$a=b$ and $\hat{a}=\hat{b}$, and accordingly 
\begin{equation}
a_{\pm}=b_{\pm}.
\label{eq:Bno32}
\end{equation}
In this way, the spectrum and polarization given in 
Eqs.~(\ref{eq:Bno06})~--~(\ref{eq:Bno15}) are simplified.

%%%%%%%%%%%%%%%%%%%%%%%%%%%%%%%%%%%%%%%%%%%%%%%%%%%%%
\subsection{Dirac neutrino case}
According to Condition (B) presented in \S\ref{sec:2no2}, the 
masses of the Dirac-type neutrinos are conjectured to be so small 
that all neutrinos are allowed to be emitted in muon decay. 
The finite neutrino masses give rise to slight deviations 
from unitarity characteristics, through the relations 
given in Eqs.~(\ref{eq:2no14}) and (\ref{eq:2no15}). 
We express these small deviations as follows:
\begin{align}
&a = 1 - \varepsilon_{a}({U^2U^2}) \Rightarrow 1,&
&b = 1 - \varepsilon_{b}({U^2U^2}) \Rightarrow a,
\label{eq:Bno49} \\
&\hat{a} = 1 - \varepsilon_{\hat{a}}({V^2V^2}) \Rightarrow 1,&
&\hat{b} = 1 - \varepsilon_{\hat{b}}({V^2V^2}) \Rightarrow \hat{a},
\label{eq:Bno51} \\
&c = 1 - \varepsilon_{c}({U^2V^2}) \Rightarrow 1,&
&\hat{c} = 1 - \varepsilon_{\hat{c}}({V^2U^2}) \Rightarrow 1.
\label{eq:Bno54}
\end{align}
Here, for example, $\varepsilon_{c}({U^2V^2})$ stands for a factor 
which is written as a product of $r_{jk}^2$ and $|U_{ej}|^2|V_{\mu k}|^2$, 
except in the narrow range near $x_{\mathrm{max}}$. 
The magnitudes of these six factors are in general less than $10^{-13}$, 
as seen from Eq.~(\ref{eq:2no09}).  
The arrows in the above expressions represent 
the limit taken under both Conditions (A) and (B) stated in \S\ref{sec:two}.  
The values of $a,\hat{a},b,\hat{b}, c$ and $\hat{c}$ can thus all be 
considered unity in practice.%
\footnote{The values of $a$ and $b$ in Table II of Ref.\,\citen{Doi2} 
should be taken as unity for the Dirac neutrino case.}

%%%%%%%%%%%%%%%%%%%%%%%%%%%%%%%%%%%%%%%%%%%%%%%%%%%%%%%
\subsection{Majorana neutrino case}
In the Majorana neutrino case, 
it is assumed that there exist neutrinos with both small 
and large masses.  
Heavy neutrinos are forbidden energetically to be emitted 
in the muon decay we consider, 
and the primed sums are therefore only over light neutrinos. 
The coefficients are expressed as follows:
\begin{align}
&a \Rightarrow \left( 1 - \overline{u_{e}}^{\, 2}\right)
\left( 1 - \overline{u_{\mu}}^{\, 2} \right), &
&b \Rightarrow a,
\label{eq:Bno57} \\
&\hat{a} \Rightarrow \,
\overline{v_{e}}^{\, 2} \,\overline{v_{\mu}}^{\, 2},&
&\hat{b} \Rightarrow \,\hat{a},
\label{eq:Bno59} \\
&c \Rightarrow 
(1 - \overline{u_{e}}^{\, 2} )\,\overline{v_{\mu}}^{\, 2},&
&\hat{c} \Rightarrow \,
\overline{v_{e}}^{\, 2} \,(1 - \overline{u_{\mu}}^{\, 2}),
\label{eq:Bno61} \\
&d \Rightarrow | \, \overline{w_{e \mu}} \, |^{2},&
&\hat{d} \Rightarrow | \, \overline{w_{e \mu \, h}}\, |^{2},
\label{eq:Bno63} \\
&d_{r} \Rightarrow
\mbox{Re} ( \overline{w_{e \mu}}^{\,\, *} \,\,
\overline{w_{e \mu \, h}} ),&
&d_{i} \Rightarrow
\mbox{Im} ( \overline{w_{e \mu}}^{\,\, *} \,\,
\overline{w_{e \mu \, h}} ).
\label{eq:Bno65}
\end{align}

The quantities $\overline{u_{\ell}}$ and $\overline{v_{\ell}}$ 
here are defined in Eq.~(\ref{eq:2no21}), 
and $\overline{w_{e \mu}}$ and $\overline{w_{e \mu \, h}}$ are 
defined in Eq.~(\ref{eq:2no23}). 
The arrows represent the limit taken under both Conditions (A) and (B) 
given in \S\ref{sec:two}. The 
quantities on the right-hand sides indicate the magnitudes of 
the coefficients in terms of the lepton mixing matrix elements.%
\footnote{The values of $a$ and $b$ listed in Table II of Ref.\,\citen{Doi2} 
should be taken as less than unity for the Majorana neutrino case.}

%%%%%%%%%%%%%%%%%%%%%%%%%%%%%%%%%%%%%%%%%%%%%%%%%%%%%%%%%%%%%%%
%%%%%%%%%%%%%%%%%         References         %%%%%%%%%%%%%%%%%%
%%%%%%%%%%%%%%%%%%%%%%%%%%%%%%%%%%%%%%%%%%%%%%%%%%%%%%%%%%%%%%%

%

\begin{thebibliography}{99}
%%%%% Chapter 1 reference  %%%%%
\bibitem{Fetscher}
For general review, see Particle Data Group (Review: W.~Fetscher 
and H.~J.~Gerber), \PLB{592,2004,410}.\\
K.~Mursula and F.~Scheck, \NPB{253,1985,189}.\\
For old articles, see the references in these papers.
%
\bibitem{TwistRho}
J.~R.~Musser et al. (TWIST Collaboration), \PRL{94,2005,101805}.
%
\bibitem{TwistDelta}
A.~Gaponenko et al. (TWIST Collaboration), \PRD{71,2005,071101}.
%
\bibitem{TransPol}
N.~Danneberg et al., \PRL{94,2005,021802}.
%
\bibitem{neutrinomass}
Y.~Fukuda et al. (Super-Kamiokande Collaboration), 
\PRL{81,1998,1562}; ibid. \textbf{86} (2000), 5651.\\
Q.~R.~Ahmad et al. (SNO Collaboration), 
\PRL{87,2001,071301}; ibid. \textbf{89} (2002), 011301.\\
K.~Eguchi et al. (KamLAND Collaboration),  \PRL{90,2003,021802}.
%
\bibitem{see-saw}
T.~Yanagida, {\it \textquotedblleft Horizontal Gauge Symmetry 
and Masses of Neutrinos\textquotedblright}, 
in {\sl Proc. Workshop on Unified Theory and Baryon Number 
in the  Universe}, eds. O. Sawada and A. Sugamoto (KEK, 1979) 95.\\ 
M.~Gell-Mann, P.~Ramond and R.~Slansky, 
{\it \textquotedblleft Complex Spinors and Unified Theories\textquotedblright}, 
in {\sl Supergravity}, 
eds. P. van Nieuwenhuizen and D. Z. Freedman (North-Holland, Amsterdam, 1979) 315.
%%
%%% Chapter 2 reference  %%%%%
%%
\bibitem{Doi1}
See, for example, 
M.~Doi, T.~Kotani and E.~Takasugi, \PTPS{83,1985,1}.
%
\bibitem{Arbuzov}
See, for QED corrections, 
A.~Arbuzov, J. High Energy Phys. {\bf 03} (2003), 063.\\
A.~B.~Arbuzov, J. High Energy Phys.\ Lett. {\bf 78} (2003), 179.\\
M.~Fischer, S.~Groote, J.~G.~K\"orner and M.~C.~Mauser, 
\PRD{67,2003,113008}.\\
For other articles, see the references in these papers.
%A.~B.~Arbuzov, \PLB{524,2002, 99}.\\
%A.~Arbuzov, A.~Czarnecki, and A.~Gaponenko, \PRD{65,2002,113006}.\\
%A.~Arbuzov and K.~Melnikov, \PRD{66,2002,093003}.
%
\bibitem{Doi2}
M.~Doi, T.~Kotani, H.~Nishiura, K.~Okuda and E.~Takasugi, 
\PTP{67,1982,281}.
%
\bibitem{Doi3}
M.~Doi, T.~Kotani, H.~Nishiura, K.~Okuda and E.~Takasugi,
Science Report (College of General Education, Osaka University) 
{\bf 30} (1981), 119.
%
\bibitem{Shrock}
R.~E.~Shrock, \PLB{112,1982,382}.
%%
%%%%% Chapter 3 reference  %%%%%
%%
%%
\bibitem{Kuno}
Y.~Kuno and Y.~Okada, Rev.\ Mod.\ Phys. {\bf 73} (2001), 151.
%%
\bibitem{Jodidio}
A.~Jodidio et al., \PRD{34,1986,1967} 
[Errata; \textbf{37} (1988), 237]. 
%%
%%
%%%%% Chapter 4 reference  %%%%%
%%
%%
\bibitem{Burkard}
H.~Burkard et al., \PLB{150,1985,242}.
%%
\bibitem{Vogel}
For general review, see Particle Data Group (Review: P.~Vogel 
and A.~Piepke), \PLB{592,2004,447}.\\
Concerning the definition of $\langle\lambda\rangle$, see 
Ref.~\citen{Doi1}.\\
For other articles, see the references in these papers.
%%
%%
%%%%% Appendix A reference  %%%%%
%%
%%
\bibitem{Doi2again}
See Ref.~\citen{Doi2} for example. 
For old articles, see the references in this paper.
%
\bibitem{Herczeg}
See, for example, 
P.~Herczeg, \PRD{34,1986,3449}.
%%
\bibitem{Bilenky}
S.~M.~Bilenky and B.~Pontecorvo, Phys.\ Rep. \textbf{41}~C (1978), 225.\\
T.~Yanagida and M.~Yoshimura, \PTP{64,1980,1870}.
%%
\bibitem{Schechter}
J.~Schechter and J.~W.~F.~Valle, \PRD{22,1980,2227}.
%%
\bibitem{Doi4}
M.~Doi, M.~Kenmoku, T.~Kotani, H.~Nishiura and E.~Takasugi, 
\PTP{70,1983,1331}.
%%
%%
%%%%% Appendix B reference  %%%%%
%%
%%
\end{thebibliography}
\end{document}